\documentclass[12pt]{article}
\usepackage[latin1]{inputenc}
\usepackage{enumerate}
\usepackage{natbib}
\usepackage{amsmath}
\usepackage{amssymb}
\usepackage{indentfirst}
\usepackage[T1]{fontenc}
\usepackage{graphicx}
\usepackage{graphics}
\usepackage{epstopdf} 
\usepackage{hyperref} %produz referência clicáveis no documento.
\usepackage{ae}
\usepackage{url}
\usepackage{caption2}
\newtheorem{theo}{Theorem}

\usepackage{vmargin}
\setmarginsrb{2.5cm}{2.cm}{2.5cm}{2.cm}{1.cm}{1mm}{1cm}{1cm}
\title{Modeling the Association Structure in Doubly Robust GEE for Longitudinal Ordinal Missing Data}
\author{José Luiz P. da Silva, Enrico A. Colosimo, Fábio N. Demarqui\\Departament of Statistics\\Universidade Federal de Minas Gerais}
\begin{document}
\maketitle
\begin{abstract}
Generalized Estimation Equations (GEE) are a well-known method for the analysis of categorical longitudinal responses. GEE method has computational simplicity and population parameter interpretation. In the presence of missing data it is only valid under the strong assumption of missing completely at random. A doubly robust estimator (DRGEE) for correlated ordinal longitudinal data is a nice approach for handling intermittently missing response and covariate under the MAR mechanism. Independent working correlation is the standard way in DRGEE. However, when covariate is not time stationary, efficiency can be gained using a structured association. The goal of this paper is to extend the DRGEE estimator to allow modeling the association structure by means of either the correlation coefficient or local odds ratio. Simulation results revealed better performance of the local odds ratio parametrization, specially for small samples. The method is applied to a data set related to Rheumatic Mitral Stenosis.
\end{abstract}

\section{Introduction}\label{dois_intro}

Longitudinal data arise when each individual is measured repeatedly through time. These repeated responses form a cluster and it is expected the response within each cluster to be correlated. A popular approach for the analysis of longitudinal data is the Generalized Estimating Equation (GEE) method, proposed by \cite{liang1986}. The goal of this procedure is to estimate fixed parameters without specifying a joint distribution for the data \citep{nooraee2014}. GEE method requires only the correct specification of the response's mean structure for the parameter estimator to be consistent and asymptotically normal. The attractive feature of GEE is that the association parameters among repeated measures are taken as a `nuisance' parameters and, unlike maximum likelihood methods, mean parameter estimates are not sensible to the specification of the association structure. Furthermore, GEE method allows marginal interpretation of the parameter of interest and it has computational simplicity.

A simple way of analyzing such correlated data is to consider an independence working assumption for the repeated responses. However, when the covariate design is not time-stationary it will lead to inefficient marginal regression estimates \citep{lipsitz1994}. In the presence of time-varying covariates efficiency can be gained assuming a different correlation structure. Nevertheless, modeling the association in ordinal data is not a simple task. Different approaches have been proposed to estimate the association parameters for ordinal responses. \cite{lipsitz1994} provided moment estimators for a variety of correlation matrices, while \cite{parsons2006} proposed an approach which estimates the correlation vector by minimizing the logarithm of the determinant of the covariance matrix of the fixed parameters. Instead of using correlations, \cite{lumley1996} proposed using a common global odds to reduce the number of association parameters. \cite{heagerty1996} extended the ALR method proposed by \cite{carey1993} to ordinal responses using a second set of estimating equations for the global odds ratio. Recently, \cite{touloumis2013} considered a family of association models to estimate local odds ratios as a measure of association. A comparison study of different working association structures can be found in \cite{nooraee2014}.

In the presence of missing data, inferences are valid if the missingness mechanism is missing completely at random (MCAR), as defined by \cite{Rubin1976}. When data is MAR, one can adopt multiple imputation GEE (MIGEE) \citep{little1987} or a weighted version (WGEE) \citep{robins1995}. These single robust versions of GEE for incomplete data require the correct specification of the weight model of GEE (WGEE) or the imputation model (MIGEE).  Doubly robust estimators (DRGEE) (\cite{carpenter2006}, \cite{tsiatis2006}, \cite{seaman2009}, \cite{chen2011}) combine ideas from these two approaches. For consistency, it requires only the weight or the imputation model to be correctly specified, providing more flexibility for the modeler.

This work was motivated by the Rheumatic Mitral Stenosis study in which a cohort of 164 patients with rheumatic mitral stenosis (a narrowing of the mitral valve in the heart) were referred for treatment at Hospital das Clinicas of the Federal University of Minas Gerais, Brazil. The response of interest was the functional classification (NYHA), major determinant of quality of life and survival of the individual. The main objective of that study was to characterize the improvement of the functional classification over time. This study was characterized by an arbitrary pattern of missing data. Response and a particular covariate (atrial compliance) were missing for some patients and the MAR mechanism seems to be a reasonable assumption for this data.

In the current paper, we consider a doubly robust approach for the analysis of longitudinal ordinal data with intermittently missing response along with a key covariate that is MAR. The cumulative logit model was used for the marginal means. We extend the doubly robust estimator to accommodate two parametrizations of the association structures: the correlation coefficient \citep{lipsitz1994} and the local odds ratio \citep{touloumis2013}. Efficiency and accuracy of the proposed estimator are compared under these association structures.

The paper is organized as follows. In Section \ref{dois_geecomp_full} are defined the notation for GEE with fully observed data and discuss the two parametrizations of the association structure. Section \ref{dois_availapp} outlines WGEE and MIGEE approaches. The proposed methodology is established in Section \ref{dois_drgee}. A simulation study is presented in Section \ref{dois_simula}, in which the finite-sample biases and mainly standard errors are compared for the standard GEE, MIGEE, WGEE and doubly robust versions, under both the correlation and local odds parametrizations. Data arising from the Rheumatic Mitral Stenosis study are analyzed in Section \ref{dois_realdata}. Paper ends with a discussion and future directions in Section \ref{dois_discuss}.

\section{Notation and GEE for Complete Data}\label{dois_geecomp_full}

In this section it is introduced the generalized estimating equations for the analysis of fully observed ordinal data. Section \ref{dois_geecomp} establishes the model and notation for longitudinal ordinal data. Section \ref{dois_covpar_est}
presents two competing ways of modeling the association structure in GEE.

\subsection{GEE for Longitudinal Ordinal Response}\label{dois_geecomp}

Let $O_{it} \in \left\{1,2,\ldots,J \right\}$ be the ordinal response for $i$-th subject $(i=1,\ldots,n)$ at time $t \ (t=1,\ldots,T_i, \ T_i\leq T)$. As the response has $J$ levels it can be defined as $Y_{itj}=I(O_{it}= j)$ for $j=1,\ldots,J$, where $I(A)$ denotes the indicator function. $Y_{itj}$ is converted into the equivalent $(J-1)$-variate vector $\boldsymbol{Y}_{it}=(Y_{it1},\ldots,Y_{it(J-1)})^T$ and let $\boldsymbol{Y}_i=(Y_{i1}^T,\ldots,Y_{iT_i}^T)^T$ be the stacked response vector. When $J=2$ the response is binary and $\boldsymbol{Y}_{it}$ is a scalar. Let $X_i$ denotes the time-stationary covariate for the $i$-th subject, and  $\boldsymbol{Z}_i=(\boldsymbol{Z}_{i1}^T,\ldots,\boldsymbol{Z}_{iT_i}^T)^T$ a $T_i\times q$ matrix of explanatory variables. 

The marginal distribution of $\boldsymbol{Y}_{it}$ is assumed to be multinomial ($\sum_{j=1}^J Y_{itj}=1$), that is
\begin{equation}
	f(\boldsymbol{Y}_{it}|\boldsymbol{X}_{it},\boldsymbol{Z}_{it},\boldsymbol{\beta})=\prod_{j=1}^J \mu_{itj}^{y_{itj}},
\end{equation}
where $\mu_{itj}=\mu_{itj}(\boldsymbol{\beta})=E(Y_{itj}|X_i,\boldsymbol{Z}_i,\boldsymbol{\beta})=Pr(O_{it}= j|X_i,\boldsymbol{Z}_i,\boldsymbol{\beta})$, is the probability of response $j$ at time $t$ and $\boldsymbol{\beta}$ is a $p \times 1$ vector of parameters. Two common choices for modeling $\mu_{itj}$ are the cumulative logit and probit models. In this work it is assumed a cumulative logit link, that is,
\begin{equation}\label{dois_mediaor}
	\mbox{logit} \left[Pr(O_{it}\leq j|X_{i},Z_{it})\right]=\beta_{0j}+X_{i}\beta_x+\boldsymbol{Z}_{it}^T\boldsymbol{\beta}_z ,\ \ \ j=1,\ldots,J-1.
\end{equation}

Formulation in (\ref{dois_mediaor}) implies a proportional odds model \citep{mccullagh1980}. In such model the interpretation of $\boldsymbol{\beta}$ is the same regardless of the number of categories (i.e., it is invariant to combination of categories). A desired feature is that the exponential of the parameters is interpreted as an odds ratio \citep{agresti2013}.

The main interest is to make inferences related to the regression parameters \\
$\boldsymbol{\beta}=(\beta_{01},\ldots,\beta_{0,J-1},\beta_x,\boldsymbol{\beta}_z^T)^T$ associated to the $(J-1)\times 1$ marginal probability vectors
\begin{equation*}
	E(Y_{it}|X_i,\boldsymbol{Z}_i)=\boldsymbol{\mu}_{it}(\boldsymbol{\beta})=(\mu_{it1},\ldots,\mu_{it(J-1)})^T.
\end{equation*}
$\boldsymbol{\mu}_{it}$ is grouped to form a vector $E(\boldsymbol{Y}_{i}|X_i,\boldsymbol{Z}_i)=\boldsymbol{\mu}_i=(\boldsymbol{\mu}_{i1}^T,\ldots,\boldsymbol{\mu}_{iT_i}^T)^T$ with the same dimension of $\boldsymbol{Y}_i$.

In order to estimate $\boldsymbol{\beta}$, generalized estimation equations are used (\cite{liang1986}; \cite{lipsitz1994}, \cite{touloumis2013}), which takes the form
\begin{equation}\label{dois_gee}
	\sum_{i=1}^n \boldsymbol{U}_i(\boldsymbol{\beta},\boldsymbol{\alpha})=\sum_{i=1}^n \boldsymbol{D}_i\boldsymbol{V}_i^{-1}(\boldsymbol{Y}_i-\boldsymbol{\mu}_i)=\boldsymbol{0},
\end{equation}
where $\boldsymbol{D}_i=\frac{\partial \boldsymbol{\mu}_i}{\partial \boldsymbol{\beta}^T}$ and $\boldsymbol{V}_i=\boldsymbol{V}_i(\boldsymbol{\beta},\boldsymbol{\alpha})$ is a $T_i(J-1)\times T_i(J-1)$ covariance matrix for $\boldsymbol{Y}_i$. The vector parameter $\boldsymbol{\alpha}$ expresses a `working' assumption about the correlation/association structure. 

Under mild regularity conditions, correct specification of the marginal mean model in $(\ref{dois_mediaor})$, and provided that a $\sqrt{n}$-consistent of $\boldsymbol{\alpha}$ is available, Liang and Zeger (1986) proved that the estimator $\boldsymbol{\hat{\beta}}$, obtained by solving $(\ref{dois_gee})$, is consistent and $\sqrt{n}(\boldsymbol{\hat{\beta}}-\boldsymbol{\beta})$ converges in distribution to a multivariate normal distribution with mean vector $\boldsymbol{0}$ and covariance matrix
\begin{equation}\label{dois_covbeta}
	\boldsymbol{\mbox{V}}_{\boldsymbol{\beta}}=\lim_{n\rightarrow \infty}n\boldsymbol{\Sigma}_0^{-1}\boldsymbol{\Sigma}_1\boldsymbol{\Sigma}_0^{-1},
\end{equation}
where $\boldsymbol{\Sigma}_0=\sum_{i=1}^n \boldsymbol{D}_i\boldsymbol{V}_i^{-1}\boldsymbol{D}_i^T$, and $\boldsymbol{\Sigma}_1=\sum_{i=1}^n\boldsymbol{D}_i\boldsymbol{V}_i^{-1}\mbox{Cov}(\boldsymbol{Y}_i)\boldsymbol{V}_i^{-1}\boldsymbol{D}_i^T$. 

In practice, the ``sandwich'' covariance matrix $\boldsymbol{\mbox{V}}_{\boldsymbol{\beta}}$ in $(\ref{dois_covbeta})$ is calculated by ignoring the limit and replacing $(\boldsymbol{\beta},\boldsymbol{\alpha})$ and $\mbox{Cov}(\boldsymbol{Y}_i)$ by $(\boldsymbol{\hat{\beta}},\boldsymbol{\hat{\alpha}})$ and $(\boldsymbol{Y}_i-\boldsymbol{\hat{\mu}}_i)(\boldsymbol{Y}_i-\boldsymbol{\hat{\mu}}_i)^T$, respectively \citep{touloumis2013}. 

\subsection{Estimation of the nuisance parameter vector and covariance matrix}\label{dois_covpar_est}

The `working' assumption is `independence' when no correlation is assumed between pairs of the response of each individual, that is, $\boldsymbol{\alpha}=\boldsymbol{0}$. Independent Estimating Equations (IEE) are proved to be efficient only when covariates are constant over time or if the independence structure is actually true \citep{lipsitz1994}. In this case, score equations of the ML method for the regression vector $\boldsymbol{\beta}$ are identical to the IEE if all observations are treated as independent. On the other hand, when there exist within-individual association or time-varying covariates, efficiency can be gained by modeling the correlation structure. Hence, a number of proposals have been formulated to estimate $\boldsymbol{\alpha}$. These alternatives differ in the efficiency of estimating the covariance matrix and computational simplicity.

\cite{lipsitz1994} defined $\boldsymbol{\alpha}$ as the correlation coefficient, and suggested the use of the method of moments to estimate a number correlation structures. \cite{parsons2006} proposed an approach that estimate the correlation vector $\boldsymbol{\alpha}$ by minimizing an objective function $Q(\boldsymbol{\alpha}|\boldsymbol{\beta},\boldsymbol{Y})$.  \cite{lumley1996} proposed using a common global odds. \cite{heagerty1996} extended the ALR method proposed by \cite{carey1993} to ordinal responses by using a second set of estimating equations for the global odds ratio. Finally, \cite{touloumis2013} identifies $\boldsymbol{\alpha}$ as a `nuisance' parameter vector that contains the marginalized local odds ratios structure. They employed a family of association models in order to develop meaningful structures for the ordinal response.

\subsubsection{Correlation Coefficient}

\cite{lipsitz1994} suggested a method that constrains the correlations at different times between two categories of the response. In their approach the weight matrix $\boldsymbol{V}_i$ is decomposed into the form $\boldsymbol{V}_i(\boldsymbol{\beta},\boldsymbol{\alpha})=\boldsymbol{F}_i^{1/2}(\boldsymbol{\beta})\boldsymbol{C}_i(\boldsymbol{\alpha})\boldsymbol{F}_i^{1/2}(\boldsymbol{\beta})$, where $\boldsymbol{F}_i$ is a matrix containing marginal variances,  $\boldsymbol{F}_{it}$, given by
\begin{equation*}
	\boldsymbol{F}_{it}=\mbox{diag} \left[\mu_{it1}(1-\mu_{it1}),\ldots,\mu_{it,J-1}(1-\mu_{it,J-1})\right],
\end{equation*}
and $\boldsymbol{C}_i$ is equal to the marginal correlation matrix. The unknown elements of $\boldsymbol{C}_i(\boldsymbol{\alpha})$ are the elements of the $(J-1)\times(J-1)$ matrix $\boldsymbol{\rho}_{itt'}(\boldsymbol{\alpha})$. Thus, $\boldsymbol{\rho}_{itt'}(\boldsymbol{\alpha})$ is parametrized as $\rho_{itt'}=Corr(Y_{it},Y_{it'})$.
The $(J-1)\times (J-1)$ diagonal blocks of $\boldsymbol{V}_i$ are $\boldsymbol{F}_{it}^{-1/2}\boldsymbol{V}_{it}\boldsymbol{F}_{it}^{-1/2}$, with $\boldsymbol{V}_{it}=\mbox{diag}(\boldsymbol{\mu}_{it})-\boldsymbol{\mu}_{it}\boldsymbol{\mu}_{it}^T$; and the $(J-1)\times (J-1)$ off-diagonal blocks of $\boldsymbol{C}_i (\boldsymbol{\alpha})$ are $\boldsymbol{\rho}_{itt'}=\boldsymbol{\rho}_{itt'}(\boldsymbol{\alpha})$, which represents the correlation between $\boldsymbol{Y}_{it}$ and $\boldsymbol{Y}_{it'}$, $t\neq t'$. The vector $\boldsymbol{\alpha}$ is a parameter vector associated with the model for $\boldsymbol{\rho}_{itt'}$. 

Define the Pearson residual vector, $\boldsymbol{e}_{it}$ as 
$$
\boldsymbol{e}_{it}=\boldsymbol{F}_{it}^{-1/2}\left( \boldsymbol{Y}_{it}-\boldsymbol{\mu}_{it} \right).
$$
Then it follows that 
$$
C_{itt'}(\alpha)=Corr(Y_{it},Y_{it'})=E(e_{it}e_{it'}^T).
$$
In order to reduce the dimension of the correlation vector, \cite{lipsitz1994} assumed an uniform correlation structure over the individuals. The use of the method of moments was suggested for a variety of correlation matrices such as
\begin{itemize}
	\item \emph{exchangeable}: $\boldsymbol{\rho}_{itt'}=\rho$, for all $t<t'$;
	\item \emph{1-dependent}: $\boldsymbol{\rho}_{it,t+1}=\rho_t$, for $t=1,\ldots,T-1$, and $\boldsymbol{\rho}_{itt'}=\boldsymbol{0}$ otherwise;
	\item \emph{banded}:  $\boldsymbol{\rho}_{itt'}=\rho_{\tau}$, when $|t'-t|=\tau$, for $\tau=1,\ldots,T-1$;
	\item \emph{unstructured}: $\boldsymbol{\rho}_{itt'}=\boldsymbol{\rho}_{itt'}$.
\end{itemize}

The estimate $\boldsymbol{\hat{\alpha}}$ is plugged into $(\ref{dois_gee})$ and a solution is found for $\boldsymbol{\beta}$. The solution might be obtained by a Fisher-scoring algorithm.

The correlation coefficient parametrization ignores the scale of the response variable and may result in loss of information regarding the correlation between the variables \citep{lumley1996}. Moreover, \cite{lipsitz1994} observed that the `working' correlation matrix is not always positive definite, which may result in a breakdown of the Fisher scoring method. This is specially true over unstructured correlations matrices and small sample sizes. When the given model for $\boldsymbol{\rho}_{itt'}$ contains too many parameters, the resulting estimates of $\boldsymbol{\beta}$ and $\boldsymbol{\alpha}$ may be highly variable.

\subsubsection{Local Odds Ratio}
Instead of modeling the correlation coefficients, some authors (see, for example, \cite{lumley1996}, \cite{touloumis2013}) modeled the off-diagonal $V_{itt'}$ through the joint probability of the responses $Y_{it}$ and $Y_{it'}$, for $t\neq t'$. The covariance of $Y_{itj}$ and $Y_{itj'}$ can be writen as
\begin{eqnarray*}
	Cov(Y_{itj},Y_{it'j'}|X_i,\boldsymbol{Z}_{it},\boldsymbol{\beta})  & = & E(Y_{itj},Y_{it'j'}|X_i,\boldsymbol{Z}_{it},\boldsymbol{\beta})-E(Y_{itj}|X_i,\boldsymbol{Z}_{it},\boldsymbol{\beta})E(Y_{it'j'}|X_i,\boldsymbol{Z}_{it},\boldsymbol{\beta})\\
	& = & \mu_{itjt'j'}-\mu_{itj}\mu_{it'j'}.
\end{eqnarray*}
The product of the first moments can be calculated through the specification of the marginal model (\ref{dois_mediaor}). The joint probabilities $\mu_{itjt'j'}=P(Y_{itj}=1,Y_{it'j'}|X_i,\boldsymbol{Z}_{it},\boldsymbol{\beta})$ can be modeled by an association vector that describes the association structure for $\forall i=1,\ldots,n$, $t\neq t'=1,\ldots,T_i$ and $j,j'=1,\ldots,J$ \citep{touloumis2013}. Estimation of these joint probabilities can be achieved through the global odds ratio (see, for example, \cite{lumley1996} and \cite{heagerty1996}) or a local odds ratio \citep{touloumis2013}. \cite{touloumis2013} argue that the local odds ratio provides the best parametrization because it is variation independent to $\boldsymbol{\beta}$ and they produce valid and unique positive joint probabilities. 

Let's introduce the local odds ratio of \cite{touloumis2013}. Denote the $L=T(T-1)/2$ time pairs as $(1,2),(1,3),\ldots,(T-1,T)$, where $T=max\{ T_1,\ldots,T_n\}$. Let $F_{tjt'j'}=\sum_{i=1}^n Y_{itj}Y_{it'j'}$ the observed frequency of the cutpoint $(j,j')$ at the $(t,t')$ time pair of the marginalized table. Define $\theta_{tjt'j'}$ as the local odds, that is, 
$$
\theta_{tjt'j'}=\frac{F_{tjt'j'}F_{t,j+1,t',j'+1}}{F_{t,j+1,t'j'}F_{tj,t',j'+1}},
$$ 
for $j,j'=1,\ldots,J-1$, and let $\boldsymbol{\alpha}$ be the $L\times(J-1)^2$ vector consisting of the local odds ratio
$$
\boldsymbol{\alpha}=(\theta_{1121},\ldots,\theta_{112(J-1)}, \ldots, \theta_{(T-1)1T1}, \ldots, \theta_{(T-1)(J-1)T(J-1)})^T.
$$

The association vector $\boldsymbol{\alpha}$ can be estimated by fitting a loglinear model for the counts $\left\{ F_{tjt'j'}\right\}$ simultaneously to all possible $L$ marginalized contingency tables and then calculating the implied local odds ratio \citep{touloumis2013}. For notational reasons, let $A$ and $B$ be the row and and column variable, respectively, and let $G$ be the group variable with levels being the $L$ ordered pairs. Assuming a Poisson sampling scheme to the $L$ sets of $J\times J$ contingency tables, fit the RC type model \citep{becker1989}
\begin{equation}\label{dois_rcmodel}
	\log (f_{tjt'j'})=\lambda+\lambda_j^A+\lambda_{j'}^B+\lambda_{(tt')}^G+\lambda_{(tt')}^{AG}+\lambda_{(tt')}^{BG}+\varphi^{(t,t')}\nu_j^{(t,t')}\nu_{j'}^{(t,t')},
\end{equation}
where $\left\{ \nu_{j}^{(t,t')}:j=1,\ldots,J \right\}$ are the score parameters for the $J$ response categories at time pair $(t,t')$, and $\{f_{tjt'j'}: j,j'=1,\ldots,J \}$ are the expected frequencies.
The maximum likelihood estimate of $\boldsymbol{\alpha}$ are obtained by treating the $L$ marginalized contingency tables as independent. By imposing identifiability constraints on the regression parameters in (\ref{dois_rcmodel}), the resulting unrestricted local odds ratio are determined by $\log(\theta_{tj't'j'})=\varphi^{(t,t')}(\nu_{j}^{(t,t')}-\nu_{j+1}^{(t,t')})(\nu_{j'}^{(t,t')}-\nu_{j'+1}^{(t,t')})$, where the intrinsic parameter $\varphi^{(t,t')}$ measures the average association of the marginalized contingency table. To increase parsimony, common unit-spaced score parameters ($\nu_{j}^{(t,t')}=j$) are usually assumed. Main options for the marginalized local odds ratio structures include
\begin{itemize}
	\item \emph{uniform}: $\log(\theta_{tj't'j'})=\varphi$, estimates a single parameter;
	\item \emph{time exchangeability}: $\log(\theta_{tj't'j'})=\varphi^{(t,t')}$; estimates $L$ parameters;
	\item \emph{category exchangeability}: $\log(\theta_{tj't'j'})=\varphi(\nu_{j}-\nu_{j+1})(\nu_{j'}-\nu_{j'+1})$; estimates $J-1$ parameters, and 
	\item \emph{unstructured}: $\log(\theta_{tj't'j'})=\varphi^{(t,t')}(\nu_{j}^{(t,t')}-\nu_{j+1}^{(t,t')})(\nu_{j'}^{(t,t')}-\nu_{j'+1}^{(t,t')})$, that requires $L(J-1)$ parameters.
\end{itemize}

Conditional on $\boldsymbol{\hat{\alpha}}$, and the marginal specification (\ref{dois_mediaor}), the joint probabilities $\mu_{itjt'j'}$ are estimated based on the adopted local odds ratio structure using the IPFP (Iterative Proportional Fitting Procedure). This algorithm, proposed by \citep{deming1940}, is used to obtain $\mu_{itjt'j'}$ through the marginals $\mu_{itj}$ and $\mu_{it'j'}$. \cite{touloumis2013} proved that the IPFP solution preserves local odds ratios of the initial values as long as they are positive. Hence, it is straightforward to calculate the weight matrix $\boldsymbol{V}_i$ and the estimating equations in (\ref{dois_gee}) can be solved with respect to $\boldsymbol{\beta}$.

An advantage of the local odds ratio parametrization over the correlation is that the local odds ratio and the marginal regression vector are variation independent. This means that $\boldsymbol{\beta}$ estimates are less sensible to a possibly wrong specification of $\boldsymbol{\alpha}$. As opposed to the correlation parametrization, the estimation of the association parameters does not depend on covariates and, as long as it is based on maximum likelihood models, the $\boldsymbol{\alpha}$ estimates are consistent under MAR. Thus, no adjustment need to be done on $\boldsymbol{\alpha}$ obtained with available data.

\section{Available Approaches for Missing Data}\label{dois_availapp}

Section \ref{dois_mecha} presents a series of assumptions related to mechanism causing data to be missing and necessary to be considered in order to build valid estimators. Multiple imputation and weighed generalized estimation equations are two commonly methods available for missing data under MAR mechanism. These methods are presented in Sections \ref{dois_migee} and \ref{dois_wgee}, respectively. They serve as the basis for the construction of the doubly robust estimator, presented in Section \ref{dois_drgee}.

\subsection{Missing Data Framework}\label{dois_mecha}

In this work it will be assumed that the time-stationary covariate $X_i$ may be missing for some subjects whereas the explanatory variables $\boldsymbol{Z}_i$ are fully observed.

For each occasion $t$ it can be defined $R_{it}=0$ if $O_{it}$ and $X_{i}$ are missing, $R_{it}=1$ if $O_{it}$ is missing and $X_{i}$ is observed, $R_{it}=2$ if $O_{it}$ is observed and $X_{i}$ is missing, and $R_{it}=3$ if $O_{it}$ and $X_{i}$ are both observed. Let $\boldsymbol{R}_i=(R_{i1},\ldots,R_{iT_i})^T$, and $\boldsymbol{\bar{R}}_{it}=(R_{i1},\ldots,R_{i,t-1})$.

The marginal probability $Pr(\boldsymbol{R}_i=\boldsymbol{r}_i|\boldsymbol{O}_i,\boldsymbol{Z}_i)$ can be obtained through conditional models of the form $Pr(R_{it}=r_{it}|\boldsymbol{\bar{R}}_{it},\boldsymbol{O}_i,\boldsymbol{Z}_i)$. This general formulation encompasses MCAR, MAR and MNAR mechanisms. In particular, the MAR mechanism requires 
\begin{equation}
	Pr(\boldsymbol{R}_{i}=\boldsymbol{r}_{i}|\boldsymbol{O}_i,\boldsymbol{Z}_i)=
	Pr(\boldsymbol{R}_{i}=\boldsymbol{r}_{i}|\boldsymbol{O}_{i}^o,\boldsymbol{Z}_{i}),
\end{equation}
where $\boldsymbol{O}_{i}^o$ denotes the observed components of $\boldsymbol{O}_{i}$. The following natural further assumption is made
\begin{equation}
	Pr(R_{it}=r_{it}|\boldsymbol{\bar{R}}_{it},\boldsymbol{O}_i,\boldsymbol{Z}_i)=
	Pr(R_{it}=r_{it}|\boldsymbol{\bar{R}}_{it},\boldsymbol{\bar{O}}_{it}^o,\boldsymbol{\bar{Z}}_{it}^o),
\end{equation}
for each time $t$, where $\boldsymbol{\bar{O}}_{it}^o$ and $\boldsymbol{\bar{Z}}_{it}^o$ are the histories of observed responses and covariates up to time $t-1$. 

Let $\pi_{it}=Pr(R_{it}=3|\boldsymbol{O}_i,\boldsymbol{Z}_i)$ be the marginal probability of observing both $\boldsymbol{O}_i$ and $X_i$ at time $t$, given the entire vectors of responses and covariates. Then, $\pi_{it}$ is expressed by
\begin{equation*}
	\pi_{it}=\sum_{r_{i1},\ldots,r_{i,t-1}}Pr(R_{it}=3,R_{i,t-1}=r_{i,t-1},\ldots,R_{i1}=r_{i1}|\boldsymbol{O}_i,\boldsymbol{Z}_i).
\end{equation*}
This marginal probability can be expressed in terms of the conditional probabilities $Pr(R_{it}=k|\boldsymbol{\bar{R}}_{it},\boldsymbol{O}_i,\boldsymbol{Z}_i)$, for $k=0,1,2,3$.
Throughout this paper it will be required the so-called \emph{positivity assumption}, that is, $\pi_{it}$ must be bounded away from zero. This condition is needed in order to guarantee the existence of $\sqrt{n}$-consistent estimators of $\boldsymbol{\beta}$ \citep{robins1995}.

\subsection{Multiple Imputation Generalized Estimating Equations}\label{dois_migee}

A imputation model commonly used to handle intermittently missing response and covariate, is imputation using chained equations (\cite{vanBuuen1999}, \cite{van2007}), which is more commonly referred to as full conditional specification (FCS). 
This approach specifies conditional distributions for each incomplete variable, conditional on all others variables in the imputation model. Starting from an initial imputation, FCS draws imputations by iterating over the conditional densities.

Denote by $\boldsymbol{\tilde{\beta}}_m$ and $\boldsymbol{\tilde{U}}_m$, respectively, the estimate of $\boldsymbol{\beta}$ and its covariance matrix from the GEE analysis of the $m$-th completed data set, $(m=1,\ldots,M)$. Following \citep{rubin1987m}, the combined point estimate for the parameter of interest $\boldsymbol{\beta}$ based on MI is simply the average of the M complete-data point estimates 
$$
\boldsymbol{\hat{\beta}}_{MI}=\frac{1}{M}\sum_{m=1}^M \boldsymbol{\tilde{\beta}}_m, 
$$
and an estimate of the covariance matrix of $\boldsymbol{\hat{\beta}}_{MI}$ is given by
$$
\boldsymbol{\widehat{U}}_{MI}=\boldsymbol{\tilde{W}}+\left(\frac{M+1}{M}\right)\boldsymbol{\tilde{B}},
$$
where
$$
\boldsymbol{\tilde{W}}=\frac{1}{M}\sum_{m=1}^M \boldsymbol{\tilde{U}}_m \ \ \ \mbox{and} \ \ \
\boldsymbol{\tilde{B}}=\frac{1}{M-1}
\sum_{m=1}^M (\boldsymbol{\tilde{\beta}}_m-\boldsymbol{\hat{\beta}_{MI}})(\boldsymbol{\tilde{\beta}}_m-\boldsymbol{\hat{\beta}_{MI}})'.
$$

\subsection{Weighted Generalized Estimating Equations}\label{dois_wgee}
\citet{robins1995} proposed a class of weighted estimating equations to allow for MAR mechanism. In binary longitudinal data, \cite{chen2011} extended the method to accommodate arbitrary patterns of both missing response and covariate. Their method was adapted here for longitudinal ordinal responses. 

Define a weight matrix $\boldsymbol{\Delta}_i=\left[\delta_{itt'} \right]_{T_i(J_i-1)\times T_i(J_i-1)},\ t=1,\ldots, T_i, t'=1,\ldots, T_i,$ where $\delta_{itt'}=\left\{I(R_{it}=1,R_{it'}=3)+I(R_{it}=3,R_{it'}=3) \right\}/\pi_{itt'}$ for $t\neq t'$, $\delta_{itt}=I(R_{it}=3)/\pi_{it}$, and $\pi_{itt'}=Pr(R_{it}=1,R_{it'}=3|\boldsymbol{O}_i,\boldsymbol{Z}_i)+Pr(R_{it}=3,R_{it'}=3|\boldsymbol{O}_i,\boldsymbol{Z}_i)$. In order to construct the weight matrix $\boldsymbol{\Delta}_i$ the conditional probabilities $Pr(R_{it}=k|\boldsymbol{\bar{R}}_{it},\boldsymbol{O}_i,\boldsymbol{Z}_i)$ are decomposed as the product of two separated logistic models. The first one models the probability of observing the potentially missing covariate $X_i$ whereas the latter models the probability of observing $Y_{it}$ conditional on observed responses and covariates up to time $t-1$.

The main idea of weighted generalized estimating equations (WGEE) lays on weighting the individual contribution to the estimating equation by introducing the weight matrix $\boldsymbol{\Delta}_i$ into the covariance matrix $\boldsymbol{V}_i$. This task can be accomplished by different ways depending on the parametrization of the association structure. The general WGEE for $\boldsymbol{\beta}$ are given by
\begin{equation}\label{wgee2}
	\sum_{i=1}^n \boldsymbol{U}_i(\boldsymbol{\beta},\boldsymbol{\alpha},\boldsymbol{\psi})=
	\sum_{i=1}^n \boldsymbol{D}_i\boldsymbol{M}_i(\boldsymbol{Y}_i-\boldsymbol{\mu}_i)=\boldsymbol{0}.
\end{equation}

When the correlation coefficient is adopted, the $\boldsymbol{M}_i$ matrix is decomposed as $\boldsymbol{M}_i=\boldsymbol{F}_i^{-1/2}(\boldsymbol{C}_i^{-1}\boldsymbol{\cdot}\boldsymbol{\Delta}_i)\boldsymbol{F}_i^{-1/2}$, where $\boldsymbol{A\cdot B}=\left[a_{it}\cdot b_{it} \right]$ denotes the Hadamard product of matrices $\boldsymbol{A}=\left[a_{it} \right]$ and $\boldsymbol{B}=\left[b_{it} \right]$. Under the local odds approach, the matrix $\boldsymbol{M}_i$ is given by  $\boldsymbol{M}_i=\boldsymbol{V}_i^{-1}\boldsymbol{\cdot}\boldsymbol{\Delta}_i$.

Conditionally on a consistent estimate of the correlation/association structure $\boldsymbol{\alpha}$, a consistent estimate for $\boldsymbol{\beta}$ can be obtained by solving (\ref{wgee2}), under the correct specification of the missing data model.

In the presence of missing data a consistent estimate for the correlation parameters can be obtained by defining a weighted observed pair of Pearson residual vector as $e_{it}^*=e_{it}(I(R_{it}=3)/\pi_{it})$. Then, a moment-based estimator can be constructed using the weighted pair contribution. For instance, an exchangeable correlation estimate can be obtained through
$$
\hat{\rho}_{itt'}=\frac{1}{\sum_{i=1}^n T_i(T_i-1) - p}\sum_{i=1}^n \sum_{t'>t} e_{it}^*e_{it'}^{*T}\frac{\pi_{it}\pi_{it'}}{\pi_{itt'}}.
$$
It is easy to show that $\hat{\rho}_{itt'}$ is an unbiased estimator.

\section{Doubly Robust GEE for Longitudinal Ordinal Data}\label{dois_drgee}

Some authors (e.g., \cite{scharfstein1999}, \cite{tsiatis2006}) noted that adding a term of expectation zero, say $\phi(\cdot)$, to the inverse probability weighted estimators would still result in consistent estimates under the MAR mechanism. The solutions of these augmented estimating equations give rise to the so-called \emph{doubly robust} estimators. 

Following \cite{chen2011}, the optimal $\phi_{opt}$ for missing response and covariate is given by $\phi_{opt}=E_{(\boldsymbol{Y}_i^m,X_i^m|\boldsymbol{Y}_i^o,X_i^o,\boldsymbol{Z}_i,\boldsymbol{R}_i)}\left\{\boldsymbol{D}_i\boldsymbol{N}_i(\boldsymbol{Y}_i-\boldsymbol{\mu}_i)\right\}$, where $\boldsymbol{Y}_i^m$ and $X_i^m$ denote the missing components of $\boldsymbol{Y}_i$ and $X_i$, respectively. When the correlation coefficient parametrization is adopted, $\boldsymbol{N}_i$ is defined as $\boldsymbol{N}_i= \boldsymbol{F}_i^{-1/2}\left\{\boldsymbol{C}_i^{-1}\boldsymbol{\cdot}(\boldsymbol{11}^T-\boldsymbol{\Delta}_i)\right\}\boldsymbol{F}_i^{-1/2}$, where $\boldsymbol{1}$ is a vector of 1's of length $T_i(J-1)$. With local odds $\boldsymbol{N}_i$ can be defined as $\boldsymbol{N}_i=\boldsymbol{V}_i^{-1}\boldsymbol{\cdot}(\boldsymbol{11}^T-\boldsymbol{\Delta}_i)$.

Conditionally on a consistent estimate of the correlation/association structure $\boldsymbol{\alpha}$, an improved estimate for $\boldsymbol{\beta}$ can then be obtained by solving the estimating equations
\begin{equation}\label{dois_dr}
	%\scriptsize
	%\sum_{i=1}^n \boldsymbol{S}_{1i}(\boldsymbol{\theta})=
	\sum_{i=1}^n \left[
	\boldsymbol{D}_i\boldsymbol{M}_i(\boldsymbol{Y}_i-\boldsymbol{\mu}_i)+
	E_{(\boldsymbol{Y}_i^m,X_i^m|\boldsymbol{Y}_i^o,X_i^o,\boldsymbol{Z}_i,\boldsymbol{R}_i)}\left\{\boldsymbol{D}_i\boldsymbol{N}_i(\boldsymbol{Y}_i-\boldsymbol{\mu}_i)\right\}
	\right]=\boldsymbol{0}.
\end{equation}
\normalsize
The estimator for $\boldsymbol{\beta}$ in (\ref{dois_dr}) is doubly-robust in the sense that it is consistent if \emph{at least one} of the missing data model or the covariate model is correctly specified \citep{chen2011}.

The referred expectation in the second part of (\ref{dois_dr}) is over the conditional distribution of $(\boldsymbol{Y}_i^m,X_i^m|\boldsymbol{Y}_i^o,X_i^o,\boldsymbol{Z}_i,\boldsymbol{R}_i)$, which can be written as
\begin{eqnarray*}
	P(\boldsymbol{Y}_i^m=\boldsymbol{y}_i^m,X_i^m=x_i^m|\boldsymbol{Y}_i^o,X_i^o,\boldsymbol{Z}_i,\boldsymbol{R}_i;\boldsymbol{\beta}^*,\boldsymbol{\gamma})
	&=&P(\boldsymbol{Y}_i^m=\boldsymbol{y}_i^m,X_i^m=x_i^m|\boldsymbol{Y}_i^o,X_i^o,\boldsymbol{Z}_i;\boldsymbol{\beta}^*,\boldsymbol{\gamma})\\
	&=&P(\boldsymbol{Y}_i^m=\boldsymbol{y}_i^m|\boldsymbol{Y}_i^o,X_i=x_i,\boldsymbol{Z}_i;\boldsymbol{\beta}^*)\\
	&&\times
	P(X_i^m=x_i^m|\boldsymbol{Y}_i^o,X_i^o,\boldsymbol{Z}_i;\boldsymbol{\gamma}).
\end{eqnarray*}

The multivariate distribution $P(\boldsymbol{Y}_i^m=\boldsymbol{y}_i^m|\boldsymbol{Y}_i^o,X_i=x_i,\boldsymbol{Z}_i;\boldsymbol{\beta}^*)$ is expressed by a product of univariate ordinal models. When $X$ is discrete, the second term in (\ref{dois_dr}) can be written as
\begin{equation*}
	E_{(\boldsymbol{Y}_i^m,X_i^m|\boldsymbol{Y}_i^o,X_i^o,\boldsymbol{Z}_i,\boldsymbol{R}_i)}\left\{\boldsymbol{D}_i\boldsymbol{N}_i(\boldsymbol{Y}_i-\boldsymbol{\mu}_i)\right\}=
	\sum_{(\boldsymbol{y}_i^m,x_i^m)}w_{ixy}
	\left\{\boldsymbol{D}_i\boldsymbol{N}_i(\boldsymbol{Y}_i-\boldsymbol{\mu}_i)\right\},
\end{equation*}
where the weight $w_{ixy}$ is given by
\begin{eqnarray*}
	w_{ixy}&=&P(\boldsymbol{Y}_i^m=\boldsymbol{y}_i^m|\boldsymbol{Y}_i^o,X_i=x_i,\boldsymbol{Z}_i;\boldsymbol{\beta}^*)\times
	P(X_i^m=x_i^m|\boldsymbol{Y}_i^o,X_i^o,\boldsymbol{Z}_i;\boldsymbol{\hat{\gamma}}).
\end{eqnarray*}
In the case of $X$ continuous, the second term in (\ref{dois_dr}) takes the form\\
\begin{equation*}
	E_{(\boldsymbol{Y}_i^m,X_i^m|\boldsymbol{Y}_i^o,X_i^o,\boldsymbol{Z}_i,\boldsymbol{R}_i)}\left\{\boldsymbol{D}_i\boldsymbol{N}_i(\boldsymbol{Y}_i-\boldsymbol{\mu}_i)\right\}=
	\int_{(\boldsymbol{Y}_i^m,X_i^m)}w_{ixy}.
	\left\{\boldsymbol{D}_i\boldsymbol{N}_i(\boldsymbol{Y}_i-\boldsymbol{\mu}_i)\right\}d\boldsymbol{Y}_i^m X_i^m,
\end{equation*}
This expectation can be cumbersome, depending on the missing data pattern. In such case, instead of using numerical integration, a Monte Carlo method can be applied to approximate the corresponding integral.

Inspired by doubly robust ideas, we constructed the following estimator for the correlation structure
$$
\hat{\rho}_{itt'}=\frac{\omega}{n^\dagger}\sum_{i=1}^n \sum_{t'>t} e_{it}^*e_{it'}^{*T}\frac{\pi_{it}\pi_{it'}}{\pi_{itt'}}
+
\frac{(1-\omega)}{n^\dagger}\sum_{i=1}^n \sum_{t'>t} \left[\sum_{(y_i^m,x_i^m)} w_{ixy} \hat{e}_{it}\hat{e}_{it'}^T\right],
$$
where $n^\dagger = T_i(T_i-1) - p$ and $0\leq \omega \leq 1$.

A sandwich estimator for the standard error of $\boldsymbol{\hat{\beta}}$ is given in Appendix.

\section{Simulation Study}\label{dois_simula}

In this section a simulation study is presented in order to investigate the performance of the proposed method under the two parametrizations of the association vector as well as its robustness to misspecification of the predictive models.  It is considered a study with $T_i=T=3$ repeated ordinal measures (with three categories) and two covariates (one quantitative and other qualitative). The true marginal model is 
\begin{equation}\label{dois_gera.resp}
	logit \ Pr(O_{it}\leq j|X_{it},Z_{it})=\beta_{0j}+\beta_1X_{i}+\beta_2Z_{it}, \ \ \ j=1,2.
\end{equation} 
where $Z_{it}\sim N(0,1/2)$ for $t=1,2,3$. 

The binary covariate $X_{it}$ may be missing for some subjects and is generated according to
\begin{equation}\label{dois_gera.cov}
	logit \ Pr(X_{i}=1|Z_{i1})=\gamma_0+\gamma_1Z_{i1}.
\end{equation}
It is assumed that $\beta_{01}=-0.4$, $\beta_{02}=1.2$, $\beta_1=-0.35$, $\beta_2=0.35$, $\gamma_0=log(1)$, $\gamma_1=2$. The correlated ordinal responses were generated using the NORTA method \citep{SimCorMultRes} with constant correlation between the latent vectors set equal to $\rho=0.7$.

In order to model $R_{it}$ two new indicators were defined. Let $R_i^x$ the indicator of observing $X_i$ and $R_{it}^y$ the indicator of observing $O_{it}$. The response variable in the first time occasion was allowed to be fully observed. The model for $R_i^x$ was defined as
\begin{equation}\label{dois_gera.perda_x}
	log\left(\frac{Pr(R_i^x=1)}{Pr(R_i^x=0)} \right)=\psi_0^x+\psi_1^xO_{i1}+\psi_2^xZ_{i1},
\end{equation}
and the model for $R_{it}^y$ was taken as
\begin{equation}\label{dois_gera.perda_y}
	log\left(\frac{Pr(R_{it}^y=1)}{Pr(R_{it}^y=0)} \right)=\psi_0^y+\psi_1^yO_{i,t-1}^*+\psi_2^yI(R_{i,t-1}^y=1)+\psi_3^yZ_{it}, \ \ t=2,3,
\end{equation}
where $O_{i,t-1}^*=O_{i,t-1}$, if $O_{i,t-1}$ is observed and $0$ otherwise. The true values are taken as $\psi_0^x=1.2$, 
$\psi_1^x=-1.5$, $\psi_2^x=-1.5$, $\psi_0^y=0.6$, $\psi_1^y=-1.5$, $\psi_2^y=2.5$, and $\psi_3^y=-1.3$. It was observed about $30\%$ of missing observations under this setup.

For comparison purposes, it was considered ordinary GEE for the complete and available data, respectively, weighted GEE (WGEE), multiple imputation (MIGEE) by chained equations \citep{mice} with $M=10$ multiple imputations, and the doubly robust versions (DRGEE). The primary goal of this simulation was to compare the performance of the above mentioned methods under the correlation and local odds ratio parametrizations. Three correlation structures (independent -- ind, exchangeable -- exch, and unstructured -- unst) and four local odds ratio structures (uniform -- unif, category exchangeability -- cat.exch, time exchangeability -- time.exch, and unstructured -- RC) were to be compared. Under independence, the estimates from the two parametrization are identical. 

In order to investigate robustness of these methods, the predicted models were also misspecified by omitting the covariate $Z_1$ from the covariate model (\ref{dois_gera.cov}) or the missing data model (\ref{dois_gera.perda_x}). 

Let $S=1000$ the total number of Monte Carlo replications. Whenever, in a given iteration, a working association structure (C) failed to converge a new sample data were generated.  Denote by $\hat{\beta}_r^C$ the corresponding GEE estimator at the $r$-th Monte Carlo replication and let $\hat{\beta}^C$ be the arithmetic mean, $\hat{\beta}^C=1/S\sum_{r=1}^S \hat{\beta}_r^C$. To evaluate the consistency of the competing methods the relative bias, defined as $100\times(\hat{\beta}^C-\beta)/\beta$, was calculated for each parameter. Interest is in quantifying the gain in efficiency by the working association structures over the independence structure. The Monte Carlo relative efficiency was defined as 
$\sum_{r=1}^S\hat{EP}(\hat{\beta_r^I})/\sum_{r=1}^S\hat{EP}(\hat{\beta_r^C})$, where $\hat{EP}(\hat{\beta_r^C})$ is the standard error of $\hat{\beta_r^C}$ based on the estimated robust covariance matrix under the (C) working association structure. The estimated coverage probability for a nominal 95\% level based on the asymptotic normality of the GEE estimators is also reported. Simulation was conducted for sample sizes $n=50, 150, 300$ and $600$. Results are presented for sample size $n=300$ subjects.

\begin{table}[htbp]
	\centering
	\scriptsize
	\caption{Evaluation criteria for misspecified models. Results for $n=300$ and $S=1000$ simulations.}\label{dois_tab1}
	\begin{tabular}{rrrrrrrrrrrrrrr}
		\hline
		& \multicolumn{4}{c}{Relative Bias} &       & \multicolumn{4}{c}{Relative Efficiency} &       & \multicolumn{4}{c}{Empirical Coverage} \\
		\cline{2-5}\cline{7-10}\cline{12-15}    Structure & $\beta_{01}$ & $\beta_{02}$ & $X$     & $Z$     &       & $\beta_{01}$ & $\beta_{02}$ & $X$     & $Z$     &       & $\beta_{01}$ & $\beta_{02}$ & $X$     & $Z$ \\
		\hline
		& \multicolumn{14}{c}{Available} \\
		ind   & \multicolumn{1}{c}{121.1} & \multicolumn{1}{c}{-31.5} & \multicolumn{1}{c}{-23.9} & \multicolumn{1}{c}{60.2} & \multicolumn{1}{c}{} & \multicolumn{1}{c}{1.00} & \multicolumn{1}{c}{1.00} & \multicolumn{1}{c}{1.00} & \multicolumn{1}{c}{1.00} & \multicolumn{1}{c}{} & \multicolumn{1}{c}{0.15} & \multicolumn{1}{c}{0.36} & \multicolumn{1}{c}{0.93} & \multicolumn{1}{c}{0.73} \\
		exch  & \multicolumn{1}{c}{94.0} & \multicolumn{1}{c}{-25.2} & \multicolumn{1}{c}{-26.4} & \multicolumn{1}{c}{32.3} & \multicolumn{1}{c}{} & \multicolumn{1}{c}{1.02} & \multicolumn{1}{c}{1.01} & \multicolumn{1}{c}{1.01} & \multicolumn{1}{c}{1.20} & \multicolumn{1}{c}{} & \multicolumn{1}{c}{0.35} & \multicolumn{1}{c}{0.55} & \multicolumn{1}{c}{0.94} & \multicolumn{1}{c}{0.86} \\
		unst  & \multicolumn{1}{c}{94.3} & \multicolumn{1}{c}{-25.9} & \multicolumn{1}{c}{-31.4} & \multicolumn{1}{c}{26.7} & \multicolumn{1}{c}{} & \multicolumn{1}{c}{1.03} & \multicolumn{1}{c}{1.01} & \multicolumn{1}{c}{1.02} & \multicolumn{1}{c}{1.19} & \multicolumn{1}{c}{} & \multicolumn{1}{c}{0.31} & \multicolumn{1}{c}{0.48} & \multicolumn{1}{c}{0.90} & \multicolumn{1}{c}{0.90} \\
		unif  & \multicolumn{1}{c}{93.7} & \multicolumn{1}{c}{-23.9} & \multicolumn{1}{c}{-27.8} & \multicolumn{1}{c}{26.4} & \multicolumn{1}{c}{} & \multicolumn{1}{c}{1.02} & \multicolumn{1}{c}{0.99} & \multicolumn{1}{c}{1.01} & \multicolumn{1}{c}{1.23} & \multicolumn{1}{c}{} & \multicolumn{1}{c}{0.33} & \multicolumn{1}{c}{0.57} & \multicolumn{1}{c}{0.92} & \multicolumn{1}{c}{0.88} \\
		cat.exch & \multicolumn{1}{c}{93.8} & \multicolumn{1}{c}{-23.9} & \multicolumn{1}{c}{-28.0} & \multicolumn{1}{c}{25.5} & \multicolumn{1}{c}{} & \multicolumn{1}{c}{1.02} & \multicolumn{1}{c}{0.99} & \multicolumn{1}{c}{1.01} & \multicolumn{1}{c}{1.24} & \multicolumn{1}{c}{} & \multicolumn{1}{c}{0.33} & \multicolumn{1}{c}{0.57} & \multicolumn{1}{c}{0.92} & \multicolumn{1}{c}{0.89} \\
		time.exch & \multicolumn{1}{c}{93.7} & \multicolumn{1}{c}{-24.0} & \multicolumn{1}{c}{-27.9} & \multicolumn{1}{c}{26.3} & \multicolumn{1}{c}{} & \multicolumn{1}{c}{1.02} & \multicolumn{1}{c}{0.99} & \multicolumn{1}{c}{1.01} & \multicolumn{1}{c}{1.24} & \multicolumn{1}{c}{} & \multicolumn{1}{c}{0.36} & \multicolumn{1}{c}{0.56} & \multicolumn{1}{c}{0.93} & \multicolumn{1}{c}{0.89} \\
		RC    & \multicolumn{1}{c}{93.1} & \multicolumn{1}{c}{-24.2} & \multicolumn{1}{c}{-30.8} & \multicolumn{1}{c}{22.5} & \multicolumn{1}{c}{} & \multicolumn{1}{c}{1.02} & \multicolumn{1}{c}{1.00} & \multicolumn{1}{c}{1.01} & \multicolumn{1}{c}{1.24} & \multicolumn{1}{c}{} & \multicolumn{1}{c}{0.34} & \multicolumn{1}{c}{0.56} & \multicolumn{1}{c}{0.92} & \multicolumn{1}{c}{0.91} \\
		& \multicolumn{14}{c}{WGEE($r^-$)} \\
		ind   & \multicolumn{1}{c}{19.9} & \multicolumn{1}{c}{-5.4} & \multicolumn{1}{c}{-28.9} & \multicolumn{1}{c}{32.8} & \multicolumn{1}{c}{} & \multicolumn{1}{c}{1.00} & \multicolumn{1}{c}{1.00} & \multicolumn{1}{c}{1.00} & \multicolumn{1}{c}{1.00} & \multicolumn{1}{c}{} & \multicolumn{1}{c}{0.94} & \multicolumn{1}{c}{0.93} & \multicolumn{1}{c}{0.92} & \multicolumn{1}{c}{0.89} \\
		exch  & \multicolumn{1}{c}{17.9} & \multicolumn{1}{c}{-5.1} & \multicolumn{1}{c}{-29.9} & \multicolumn{1}{c}{22.2} & \multicolumn{1}{c}{} & \multicolumn{1}{c}{1.00} & \multicolumn{1}{c}{0.99} & \multicolumn{1}{c}{1.00} & \multicolumn{1}{c}{1.17} & \multicolumn{1}{c}{} & \multicolumn{1}{c}{0.94} & \multicolumn{1}{c}{0.94} & \multicolumn{1}{c}{0.94} & \multicolumn{1}{c}{0.91} \\
		unst  & \multicolumn{1}{c}{21.0} & \multicolumn{1}{c}{-6.5} & \multicolumn{1}{c}{-35.0} & \multicolumn{1}{c}{15.2} & \multicolumn{1}{c}{} & \multicolumn{1}{c}{1.01} & \multicolumn{1}{c}{0.99} & \multicolumn{1}{c}{1.01} & \multicolumn{1}{c}{1.16} & \multicolumn{1}{c}{} & \multicolumn{1}{c}{0.93} & \multicolumn{1}{c}{0.92} & \multicolumn{1}{c}{0.91} & \multicolumn{1}{c}{0.94} \\
		unif  & \multicolumn{1}{c}{18.8} & \multicolumn{1}{c}{-5.1} & \multicolumn{1}{c}{-28.6} & \multicolumn{1}{c}{22.7} & \multicolumn{1}{c}{} & \multicolumn{1}{c}{1.00} & \multicolumn{1}{c}{0.99} & \multicolumn{1}{c}{1.00} & \multicolumn{1}{c}{1.16} & \multicolumn{1}{c}{} & \multicolumn{1}{c}{0.95} & \multicolumn{1}{c}{0.94} & \multicolumn{1}{c}{0.92} & \multicolumn{1}{c}{0.92} \\
		cat.exch & \multicolumn{1}{c}{19.0} & \multicolumn{1}{c}{-5.2} & \multicolumn{1}{c}{-28.8} & \multicolumn{1}{c}{22.0} & \multicolumn{1}{c}{} & \multicolumn{1}{c}{1.00} & \multicolumn{1}{c}{0.99} & \multicolumn{1}{c}{1.00} & \multicolumn{1}{c}{1.17} & \multicolumn{1}{c}{} & \multicolumn{1}{c}{0.95} & \multicolumn{1}{c}{0.94} & \multicolumn{1}{c}{0.92} & \multicolumn{1}{c}{0.92} \\
		time.exch & \multicolumn{1}{c}{18.3} & \multicolumn{1}{c}{-5.0} & \multicolumn{1}{c}{-27.7} & \multicolumn{1}{c}{22.8} & \multicolumn{1}{c}{} & \multicolumn{1}{c}{1.00} & \multicolumn{1}{c}{0.99} & \multicolumn{1}{c}{1.00} & \multicolumn{1}{c}{1.17} & \multicolumn{1}{c}{} & \multicolumn{1}{c}{0.95} & \multicolumn{1}{c}{0.93} & \multicolumn{1}{c}{0.93} & \multicolumn{1}{c}{0.92} \\
		RC    & \multicolumn{1}{c}{18.7} & \multicolumn{1}{c}{-5.6} & \multicolumn{1}{c}{-31.4} & \multicolumn{1}{c}{19.8} & \multicolumn{1}{c}{} & \multicolumn{1}{c}{1.01} & \multicolumn{1}{c}{0.99} & \multicolumn{1}{c}{1.00} & \multicolumn{1}{c}{1.17} & \multicolumn{1}{c}{} & \multicolumn{1}{c}{0.94} & \multicolumn{1}{c}{0.93} & \multicolumn{1}{c}{0.92} & \multicolumn{1}{c}{0.91} \\
		& \multicolumn{14}{c}{MIGEE($x^-$)} \\
		ind   & \multicolumn{1}{c}{20.9} & \multicolumn{1}{c}{-7.1} & \multicolumn{1}{c}{-48.0} & \multicolumn{1}{c}{-8.3} & \multicolumn{1}{c}{} & \multicolumn{1}{c}{1.00} & \multicolumn{1}{c}{1.00} & \multicolumn{1}{c}{1.00} & \multicolumn{1}{c}{1.00} & \multicolumn{1}{c}{} & \multicolumn{1}{c}{0.93} & \multicolumn{1}{c}{0.93} & \multicolumn{1}{c}{0.92} & \multicolumn{1}{c}{0.94} \\
		exch  & \multicolumn{1}{c}{19.9} & \multicolumn{1}{c}{-6.8} & \multicolumn{1}{c}{-47.9} & \multicolumn{1}{c}{-1.6} & \multicolumn{1}{c}{} & \multicolumn{1}{c}{0.99} & \multicolumn{1}{c}{0.99} & \multicolumn{1}{c}{1.00} & \multicolumn{1}{c}{1.22} & \multicolumn{1}{c}{} & \multicolumn{1}{c}{0.94} & \multicolumn{1}{c}{0.94} & \multicolumn{1}{c}{0.93} & \multicolumn{1}{c}{0.95} \\
		unst  & \multicolumn{1}{c}{21.5} & \multicolumn{1}{c}{-7.6} & \multicolumn{1}{c}{-51.9} & \multicolumn{1}{c}{-5.9} & \multicolumn{1}{c}{} & \multicolumn{1}{c}{0.99} & \multicolumn{1}{c}{1.00} & \multicolumn{1}{c}{1.00} & \multicolumn{1}{c}{1.22} & \multicolumn{1}{c}{} & \multicolumn{1}{c}{0.94} & \multicolumn{1}{c}{0.93} & \multicolumn{1}{c}{0.92} & \multicolumn{1}{c}{0.94} \\
		unif  & \multicolumn{1}{c}{19.7} & \multicolumn{1}{c}{-6.7} & \multicolumn{1}{c}{-46.8} & \multicolumn{1}{c}{-2.2} & \multicolumn{1}{c}{} & \multicolumn{1}{c}{1.00} & \multicolumn{1}{c}{1.00} & \multicolumn{1}{c}{1.00} & \multicolumn{1}{c}{1.22} & \multicolumn{1}{c}{} & \multicolumn{1}{c}{0.95} & \multicolumn{1}{c}{0.94} & \multicolumn{1}{c}{0.93} & \multicolumn{1}{c}{0.94} \\
		cat.exch & \multicolumn{1}{c}{19.9} & \multicolumn{1}{c}{-6.8} & \multicolumn{1}{c}{-47.0} & \multicolumn{1}{c}{-3.1} & \multicolumn{1}{c}{} & \multicolumn{1}{c}{1.00} & \multicolumn{1}{c}{1.00} & \multicolumn{1}{c}{1.00} & \multicolumn{1}{c}{1.23} & \multicolumn{1}{c}{} & \multicolumn{1}{c}{0.95} & \multicolumn{1}{c}{0.94} & \multicolumn{1}{c}{0.93} & \multicolumn{1}{c}{0.94} \\
		time.exch & \multicolumn{1}{c}{20.0} & \multicolumn{1}{c}{-6.8} & \multicolumn{1}{c}{-46.6} & \multicolumn{1}{c}{-2.8} & \multicolumn{1}{c}{} & \multicolumn{1}{c}{1.00} & \multicolumn{1}{c}{1.00} & \multicolumn{1}{c}{1.00} & \multicolumn{1}{c}{1.23} & \multicolumn{1}{c}{} & \multicolumn{1}{c}{0.94} & \multicolumn{1}{c}{0.94} & \multicolumn{1}{c}{0.94} & \multicolumn{1}{c}{0.94} \\
		RC    & \multicolumn{1}{c}{20.1} & \multicolumn{1}{c}{-7.3} & \multicolumn{1}{c}{-49.7} & \multicolumn{1}{c}{-5.8} & \multicolumn{1}{c}{} & \multicolumn{1}{c}{1.00} & \multicolumn{1}{c}{1.01} & \multicolumn{1}{c}{1.00} & \multicolumn{1}{c}{1.23} & \multicolumn{1}{c}{} & \multicolumn{1}{c}{0.94} & \multicolumn{1}{c}{0.93} & \multicolumn{1}{c}{0.92} & \multicolumn{1}{c}{0.93} \\
		& \multicolumn{14}{c}{DRGEE($x^-,r^-$)} \\
		ind   & \multicolumn{1}{c}{17.0} & \multicolumn{1}{c}{-5.5} & \multicolumn{1}{c}{-41.4} & \multicolumn{1}{c}{-8.9} & \multicolumn{1}{c}{} & \multicolumn{1}{c}{1.00} & \multicolumn{1}{c}{1.00} & \multicolumn{1}{c}{1.00} & \multicolumn{1}{c}{1.00} & \multicolumn{1}{c}{} & \multicolumn{1}{c}{0.93} & \multicolumn{1}{c}{0.92} & \multicolumn{1}{c}{0.89} & \multicolumn{1}{c}{0.94} \\
		exch  & \multicolumn{1}{c}{15.7} & \multicolumn{1}{c}{-5.1} & \multicolumn{1}{c}{-40.7} & \multicolumn{1}{c}{-2.4} & \multicolumn{1}{c}{} & \multicolumn{1}{c}{0.99} & \multicolumn{1}{c}{0.99} & \multicolumn{1}{c}{1.00} & \multicolumn{1}{c}{1.14} & \multicolumn{1}{c}{} & \multicolumn{1}{c}{0.94} & \multicolumn{1}{c}{0.94} & \multicolumn{1}{c}{0.91} & \multicolumn{1}{c}{0.94} \\
		unst  & \multicolumn{1}{c}{18.5} & \multicolumn{1}{c}{-6.3} & \multicolumn{1}{c}{-45.5} & \multicolumn{1}{c}{-8.5} & \multicolumn{1}{c}{} & \multicolumn{1}{c}{1.00} & \multicolumn{1}{c}{1.00} & \multicolumn{1}{c}{1.01} & \multicolumn{1}{c}{1.14} & \multicolumn{1}{c}{} & \multicolumn{1}{c}{0.93} & \multicolumn{1}{c}{0.92} & \multicolumn{1}{c}{0.89} & \multicolumn{1}{c}{0.95} \\
		unif  & \multicolumn{1}{c}{15.5} & \multicolumn{1}{c}{-5.0} & \multicolumn{1}{c}{-39.1} & \multicolumn{1}{c}{-3.1} & \multicolumn{1}{c}{} & \multicolumn{1}{c}{1.00} & \multicolumn{1}{c}{0.99} & \multicolumn{1}{c}{1.00} & \multicolumn{1}{c}{1.14} & \multicolumn{1}{c}{} & \multicolumn{1}{c}{0.94} & \multicolumn{1}{c}{0.94} & \multicolumn{1}{c}{0.91} & \multicolumn{1}{c}{0.94} \\
		cat.exch & \multicolumn{1}{c}{15.6} & \multicolumn{1}{c}{-5.1} & \multicolumn{1}{c}{-39.3} & \multicolumn{1}{c}{-3.5} & \multicolumn{1}{c}{} & \multicolumn{1}{c}{1.00} & \multicolumn{1}{c}{0.99} & \multicolumn{1}{c}{1.00} & \multicolumn{1}{c}{1.15} & \multicolumn{1}{c}{} & \multicolumn{1}{c}{0.95} & \multicolumn{1}{c}{0.94} & \multicolumn{1}{c}{0.91} & \multicolumn{1}{c}{0.94} \\
		time.exch & \multicolumn{1}{c}{15.0} & \multicolumn{1}{c}{-4.8} & \multicolumn{1}{c}{-38.1} & \multicolumn{1}{c}{-2.5} & \multicolumn{1}{c}{} & \multicolumn{1}{c}{1.00} & \multicolumn{1}{c}{0.99} & \multicolumn{1}{c}{1.00} & \multicolumn{1}{c}{1.14} & \multicolumn{1}{c}{} & \multicolumn{1}{c}{0.94} & \multicolumn{1}{c}{0.92} & \multicolumn{1}{c}{0.91} & \multicolumn{1}{c}{0.94} \\
		RC    & \multicolumn{1}{c}{15.5} & \multicolumn{1}{c}{-5.5} & \multicolumn{1}{c}{-41.1} & \multicolumn{1}{c}{-5.2} & \multicolumn{1}{c}{} & \multicolumn{1}{c}{1.00} & \multicolumn{1}{c}{0.99} & \multicolumn{1}{c}{1.00} & \multicolumn{1}{c}{1.14} & \multicolumn{1}{c}{} & \multicolumn{1}{c}{0.94} & \multicolumn{1}{c}{0.93} & \multicolumn{1}{c}{0.91} & \multicolumn{1}{c}{0.93}\\
		\hline
		\multicolumn{14}{c}{``$^+$" indicates correctly specified model and ``$^-$" indicates misspecified model omitting the $Z_1$ predictor}
	\end{tabular}%
\end{table}%

Table \ref{dois_tab1} presents the simulation results associated with ordinary GEE for available data and incorrectly specified methods, those in which the covariate $Z_1$ was omitted from their predictive models. For each method the first three lines refer to correlation structures and the last four ones refer to local odds ratio structures. These seven structures are distinguished in terms of the number of parameters being estimated as well as the restrictions placed on associations/correlations between the response indicators at different time pairs. 

Regarding the GEE for available data, the missing data impact on bias can be noticed for all parameters, the higher biases being observed in the first intercept, followed by the parameter associated with the covariate $Z$. The impact of the bias on parameter estimates was also clearly noticed by the low coverage rates. Still considering the available data it is worth noting that the bias for the parameter associated with covariate $Z$ was reduced by more than half when the association structure is modeled. All methods being compared exhibited bias when their predictive models are incorrectly specified, although the bias for the two intercepts was considerably reduced. For the parameters associated with the covariates, the bias of WGEE method was the same magnitude as those provided by ordinary GEE, and the performance of DRGEE was slightly superior to MIGEE. 

Independent estimating equations are efficient for the intercept parameters and regression coefficient associated with the baseline covariate $X$. As expected, the gain in efficiency by modeling the association structure occurs only for the time-varying covariate $Z$. Comparing to the independence structure, the gain in efficiency ranged from 14\% on average for DRGEE, 23\% for multiple imputation and about 17\% for WGEE. 

All misspecified methods presented empirical coverage rates below nominal level although they are somewhat close to the expected value in some cases, particularly for MIGEE.

\begin{table}[htbp]
	\centering
	\scriptsize
	\caption{Evaluation criteria for correctly specified models. Results for $n=300$ and $S=1000$ simulations.}\label{dois_tab2}
	\begin{tabular}{rrrrrrrrrrrrrrr}
		\hline
		& \multicolumn{4}{c}{Relative Bias} &       & \multicolumn{4}{c}{Relative Efficiency} &       & \multicolumn{4}{c}{Empirical Coverage} \\
		\cline{2-5}\cline{7-10}\cline{12-15}    Structure & $\beta_{01}$ & $\beta_{02}$ & $X$     & $Z$     &       & $\beta_{01}$ & $\beta_{02}$ & $X$     & $Z$     &       & $\beta_{01}$ & $\beta_{02}$ & $X$     & $Z$ \\
		\hline
		& \multicolumn{14}{c}{Complete} \\
		ind   & \multicolumn{1}{c}{2.1} & \multicolumn{1}{c}{0.1} & \multicolumn{1}{c}{-1.1} & \multicolumn{1}{c}{-0.5} & \multicolumn{1}{c}{} & \multicolumn{1}{c}{1.00} & \multicolumn{1}{c}{1.00} & \multicolumn{1}{c}{1.00} & \multicolumn{1}{c}{1.00} & \multicolumn{1}{c}{} & \multicolumn{1}{c}{0.96} & \multicolumn{1}{c}{0.94} & \multicolumn{1}{c}{0.95} & \multicolumn{1}{c}{0.95} \\
		exch  & \multicolumn{1}{c}{0.3} & \multicolumn{1}{c}{0.6} & \multicolumn{1}{c}{-0.2} & \multicolumn{1}{c}{1.3} & \multicolumn{1}{c}{} & \multicolumn{1}{c}{0.99} & \multicolumn{1}{c}{0.99} & \multicolumn{1}{c}{0.99} & \multicolumn{1}{c}{1.25} & \multicolumn{1}{c}{} & \multicolumn{1}{c}{0.96} & \multicolumn{1}{c}{0.95} & \multicolumn{1}{c}{0.97} & \multicolumn{1}{c}{0.96} \\
		unst  & \multicolumn{1}{c}{0.9} & \multicolumn{1}{c}{0.0} & \multicolumn{1}{c}{-3.4} & \multicolumn{1}{c}{-1.7} & \multicolumn{1}{c}{} & \multicolumn{1}{c}{1.00} & \multicolumn{1}{c}{1.00} & \multicolumn{1}{c}{1.01} & \multicolumn{1}{c}{1.25} & \multicolumn{1}{c}{} & \multicolumn{1}{c}{0.95} & \multicolumn{1}{c}{0.94} & \multicolumn{1}{c}{0.95} & \multicolumn{1}{c}{0.96} \\
		unif  & \multicolumn{1}{c}{0.1} & \multicolumn{1}{c}{0.7} & \multicolumn{1}{c}{0.7} & \multicolumn{1}{c}{0.3} & \multicolumn{1}{c}{} & \multicolumn{1}{c}{1.00} & \multicolumn{1}{c}{1.00} & \multicolumn{1}{c}{1.00} & \multicolumn{1}{c}{1.27} & \multicolumn{1}{c}{} & \multicolumn{1}{c}{0.96} & \multicolumn{1}{c}{0.95} & \multicolumn{1}{c}{0.94} & \multicolumn{1}{c}{0.94} \\
		cat.exch & \multicolumn{1}{c}{0.2} & \multicolumn{1}{c}{0.7} & \multicolumn{1}{c}{0.6} & \multicolumn{1}{c}{-0.2} & \multicolumn{1}{c}{} & \multicolumn{1}{c}{1.00} & \multicolumn{1}{c}{1.00} & \multicolumn{1}{c}{1.00} & \multicolumn{1}{c}{1.27} & \multicolumn{1}{c}{} & \multicolumn{1}{c}{0.96} & \multicolumn{1}{c}{0.95} & \multicolumn{1}{c}{0.95} & \multicolumn{1}{c}{0.94} \\
		time.exch & \multicolumn{1}{c}{0.2} & \multicolumn{1}{c}{0.3} & \multicolumn{1}{c}{0.1} & \multicolumn{1}{c}{0.3} & \multicolumn{1}{c}{} & \multicolumn{1}{c}{1.00} & \multicolumn{1}{c}{1.00} & \multicolumn{1}{c}{1.00} & \multicolumn{1}{c}{1.27} & \multicolumn{1}{c}{} & \multicolumn{1}{c}{0.95} & \multicolumn{1}{c}{0.95} & \multicolumn{1}{c}{0.96} & \multicolumn{1}{c}{0.94} \\
		RC    & \multicolumn{1}{c}{0.2} & \multicolumn{1}{c}{0.3} & \multicolumn{1}{c}{-0.8} & \multicolumn{1}{c}{-1.9} & \multicolumn{1}{c}{} & \multicolumn{1}{c}{1.00} & \multicolumn{1}{c}{1.00} & \multicolumn{1}{c}{1.00} & \multicolumn{1}{c}{1.27} & \multicolumn{1}{c}{} & \multicolumn{1}{c}{0.95} & \multicolumn{1}{c}{0.94} & \multicolumn{1}{c}{0.95} & \multicolumn{1}{c}{0.94} \\
		& \multicolumn{14}{c}{WGEE($r^+$)} \\
		ind   & \multicolumn{1}{c}{3.6} & \multicolumn{1}{c}{-0.4} & \multicolumn{1}{c}{-1.8} & \multicolumn{1}{c}{-0.7} & \multicolumn{1}{c}{} & \multicolumn{1}{c}{1.00} & \multicolumn{1}{c}{1.00} & \multicolumn{1}{c}{1.00} & \multicolumn{1}{c}{1.00} & \multicolumn{1}{c}{} & \multicolumn{1}{c}{0.96} & \multicolumn{1}{c}{0.95} & \multicolumn{1}{c}{0.94} & \multicolumn{1}{c}{0.96} \\
		exch  & \multicolumn{1}{c}{1.3} & \multicolumn{1}{c}{0.3} & \multicolumn{1}{c}{0.2} & \multicolumn{1}{c}{0.6} & \multicolumn{1}{c}{} & \multicolumn{1}{c}{0.99} & \multicolumn{1}{c}{0.98} & \multicolumn{1}{c}{1.00} & \multicolumn{1}{c}{1.19} & \multicolumn{1}{c}{} & \multicolumn{1}{c}{0.96} & \multicolumn{1}{c}{0.96} & \multicolumn{1}{c}{0.96} & \multicolumn{1}{c}{0.96} \\
		unst  & \multicolumn{1}{c}{6.2} & \multicolumn{1}{c}{-1.6} & \multicolumn{1}{c}{-6.8} & \multicolumn{1}{c}{-4.1} & \multicolumn{1}{c}{} & \multicolumn{1}{c}{1.00} & \multicolumn{1}{c}{1.00} & \multicolumn{1}{c}{1.01} & \multicolumn{1}{c}{1.19} & \multicolumn{1}{c}{} & \multicolumn{1}{c}{0.95} & \multicolumn{1}{c}{0.94} & \multicolumn{1}{c}{0.93} & \multicolumn{1}{c}{0.96} \\
		unif  & \multicolumn{1}{c}{2.4} & \multicolumn{1}{c}{-0.1} & \multicolumn{1}{c}{-0.3} & \multicolumn{1}{c}{-0.4} & \multicolumn{1}{c}{} & \multicolumn{1}{c}{0.99} & \multicolumn{1}{c}{0.99} & \multicolumn{1}{c}{1.00} & \multicolumn{1}{c}{1.18} & \multicolumn{1}{c}{} & \multicolumn{1}{c}{0.97} & \multicolumn{1}{c}{0.96} & \multicolumn{1}{c}{0.96} & \multicolumn{1}{c}{0.95} \\
		cat.exch & \multicolumn{1}{c}{2.6} & \multicolumn{1}{c}{-0.2} & \multicolumn{1}{c}{-0.5} & \multicolumn{1}{c}{-1.0} & \multicolumn{1}{c}{} & \multicolumn{1}{c}{0.99} & \multicolumn{1}{c}{0.99} & \multicolumn{1}{c}{1.00} & \multicolumn{1}{c}{1.18} & \multicolumn{1}{c}{} & \multicolumn{1}{c}{0.97} & \multicolumn{1}{c}{0.95} & \multicolumn{1}{c}{0.96} & \multicolumn{1}{c}{0.95} \\
		time.exch & \multicolumn{1}{c}{1.4} & \multicolumn{1}{c}{0.3} & \multicolumn{1}{c}{1.8} & \multicolumn{1}{c}{1.1} & \multicolumn{1}{c}{} & \multicolumn{1}{c}{0.99} & \multicolumn{1}{c}{0.99} & \multicolumn{1}{c}{1.00} & \multicolumn{1}{c}{1.19} & \multicolumn{1}{c}{} & \multicolumn{1}{c}{0.97} & \multicolumn{1}{c}{0.97} & \multicolumn{1}{c}{0.96} & \multicolumn{1}{c}{0.96} \\
		RC    & \multicolumn{1}{c}{2.5} & \multicolumn{1}{c}{-0.5} & \multicolumn{1}{c}{-2.5} & \multicolumn{1}{c}{-1.6} & \multicolumn{1}{c}{} & \multicolumn{1}{c}{0.99} & \multicolumn{1}{c}{0.99} & \multicolumn{1}{c}{1.00} & \multicolumn{1}{c}{1.19} & \multicolumn{1}{c}{} & \multicolumn{1}{c}{0.96} & \multicolumn{1}{c}{0.96} & \multicolumn{1}{c}{0.95} & \multicolumn{1}{c}{0.94} \\
		& \multicolumn{14}{c}{MIGEE($x^+$)} \\
		ind   & \multicolumn{1}{c}{5.3} & \multicolumn{1}{c}{-1.3} & \multicolumn{1}{c}{-7.1} & \multicolumn{1}{c}{-0.4} & \multicolumn{1}{c}{} & \multicolumn{1}{c}{1.00} & \multicolumn{1}{c}{1.00} & \multicolumn{1}{c}{1.00} & \multicolumn{1}{c}{1.00} & \multicolumn{1}{c}{} & \multicolumn{1}{c}{0.96} & \multicolumn{1}{c}{0.94} & \multicolumn{1}{c}{0.94} & \multicolumn{1}{c}{0.94} \\
		exch  & \multicolumn{1}{c}{3.5} & \multicolumn{1}{c}{-0.8} & \multicolumn{1}{c}{-5.6} & \multicolumn{1}{c}{0.7} & \multicolumn{1}{c}{} & \multicolumn{1}{c}{0.99} & \multicolumn{1}{c}{0.99} & \multicolumn{1}{c}{1.00} & \multicolumn{1}{c}{1.22} & \multicolumn{1}{c}{} & \multicolumn{1}{c}{0.95} & \multicolumn{1}{c}{0.95} & \multicolumn{1}{c}{0.96} & \multicolumn{1}{c}{0.95} \\
		unst  & \multicolumn{1}{c}{4.6} & \multicolumn{1}{c}{-1.5} & \multicolumn{1}{c}{-9.4} & \multicolumn{1}{c}{-3.7} & \multicolumn{1}{c}{} & \multicolumn{1}{c}{0.99} & \multicolumn{1}{c}{0.99} & \multicolumn{1}{c}{1.00} & \multicolumn{1}{c}{1.21} & \multicolumn{1}{c}{} & \multicolumn{1}{c}{0.95} & \multicolumn{1}{c}{0.95} & \multicolumn{1}{c}{0.95} & \multicolumn{1}{c}{0.94} \\
		unif  & \multicolumn{1}{c}{3.9} & \multicolumn{1}{c}{-0.8} & \multicolumn{1}{c}{-5.4} & \multicolumn{1}{c}{0.2} & \multicolumn{1}{c}{} & \multicolumn{1}{c}{1.00} & \multicolumn{1}{c}{0.99} & \multicolumn{1}{c}{1.00} & \multicolumn{1}{c}{1.22} & \multicolumn{1}{c}{} & \multicolumn{1}{c}{0.96} & \multicolumn{1}{c}{0.95} & \multicolumn{1}{c}{0.95} & \multicolumn{1}{c}{0.94} \\
		cat.exch & \multicolumn{1}{c}{4.2} & \multicolumn{1}{c}{-0.9} & \multicolumn{1}{c}{-5.7} & \multicolumn{1}{c}{-0.7} & \multicolumn{1}{c}{} & \multicolumn{1}{c}{1.00} & \multicolumn{1}{c}{0.99} & \multicolumn{1}{c}{1.00} & \multicolumn{1}{c}{1.23} & \multicolumn{1}{c}{} & \multicolumn{1}{c}{0.96} & \multicolumn{1}{c}{0.95} & \multicolumn{1}{c}{0.96} & \multicolumn{1}{c}{0.94} \\
		time.exch & \multicolumn{1}{c}{4.2} & \multicolumn{1}{c}{-1.0} & \multicolumn{1}{c}{-5.2} & \multicolumn{1}{c}{-0.2} & \multicolumn{1}{c}{} & \multicolumn{1}{c}{1.00} & \multicolumn{1}{c}{0.99} & \multicolumn{1}{c}{1.00} & \multicolumn{1}{c}{1.22} & \multicolumn{1}{c}{} & \multicolumn{1}{c}{0.95} & \multicolumn{1}{c}{0.96} & \multicolumn{1}{c}{0.96} & \multicolumn{1}{c}{0.94} \\
		RC    & \multicolumn{1}{c}{3.7} & \multicolumn{1}{c}{-1.3} & \multicolumn{1}{c}{-8.0} & \multicolumn{1}{c}{-3.1} & \multicolumn{1}{c}{} & \multicolumn{1}{c}{1.00} & \multicolumn{1}{c}{1.00} & \multicolumn{1}{c}{1.00} & \multicolumn{1}{c}{1.23} & \multicolumn{1}{c}{} & \multicolumn{1}{c}{0.94} & \multicolumn{1}{c}{0.95} & \multicolumn{1}{c}{0.95} & \multicolumn{1}{c}{0.92} \\
		& \multicolumn{14}{c}{DRGEE($x^+,r^+$)} \\
		ind   & \multicolumn{1}{c}{1.0} & \multicolumn{1}{c}{0.4} & \multicolumn{1}{c}{-1.7} & \multicolumn{1}{c}{-1.4} & \multicolumn{1}{c}{} & \multicolumn{1}{c}{1.00} & \multicolumn{1}{c}{1.00} & \multicolumn{1}{c}{1.00} & \multicolumn{1}{c}{1.00} & \multicolumn{1}{c}{} & \multicolumn{1}{c}{0.95} & \multicolumn{1}{c}{0.95} & \multicolumn{1}{c}{0.93} & \multicolumn{1}{c}{0.95} \\
		exch  & \multicolumn{1}{c}{-0.8} & \multicolumn{1}{c}{0.9} & \multicolumn{1}{c}{0.3} & \multicolumn{1}{c}{1.4} & \multicolumn{1}{c}{} & \multicolumn{1}{c}{0.99} & \multicolumn{1}{c}{0.99} & \multicolumn{1}{c}{1.00} & \multicolumn{1}{c}{1.16} & \multicolumn{1}{c}{} & \multicolumn{1}{c}{0.95} & \multicolumn{1}{c}{0.95} & \multicolumn{1}{c}{0.96} & \multicolumn{1}{c}{0.95} \\
		unst  & \multicolumn{1}{c}{2.4} & \multicolumn{1}{c}{-0.5} & \multicolumn{1}{c}{-5.7} & \multicolumn{1}{c}{-3.6} & \multicolumn{1}{c}{} & \multicolumn{1}{c}{1.00} & \multicolumn{1}{c}{0.99} & \multicolumn{1}{c}{1.01} & \multicolumn{1}{c}{1.16} & \multicolumn{1}{c}{} & \multicolumn{1}{c}{0.94} & \multicolumn{1}{c}{0.94} & \multicolumn{1}{c}{0.94} & \multicolumn{1}{c}{0.94} \\
		unif  & \multicolumn{1}{c}{0.0} & \multicolumn{1}{c}{0.7} & \multicolumn{1}{c}{-0.6} & \multicolumn{1}{c}{-0.1} & \multicolumn{1}{c}{} & \multicolumn{1}{c}{1.00} & \multicolumn{1}{c}{0.99} & \multicolumn{1}{c}{1.00} & \multicolumn{1}{c}{1.16} & \multicolumn{1}{c}{} & \multicolumn{1}{c}{0.95} & \multicolumn{1}{c}{0.95} & \multicolumn{1}{c}{0.95} & \multicolumn{1}{c}{0.94} \\
		cat.exch & \multicolumn{1}{c}{0.1} & \multicolumn{1}{c}{0.6} & \multicolumn{1}{c}{-0.8} & \multicolumn{1}{c}{-0.6} & \multicolumn{1}{c}{} & \multicolumn{1}{c}{1.00} & \multicolumn{1}{c}{0.99} & \multicolumn{1}{c}{1.00} & \multicolumn{1}{c}{1.17} & \multicolumn{1}{c}{} & \multicolumn{1}{c}{0.95} & \multicolumn{1}{c}{0.95} & \multicolumn{1}{c}{0.95} & \multicolumn{1}{c}{0.94} \\
		time.exch & \multicolumn{1}{c}{-0.9} & \multicolumn{1}{c}{1.2} & \multicolumn{1}{c}{2.0} & \multicolumn{1}{c}{1.4} & \multicolumn{1}{c}{} & \multicolumn{1}{c}{0.99} & \multicolumn{1}{c}{0.99} & \multicolumn{1}{c}{1.00} & \multicolumn{1}{c}{1.16} & \multicolumn{1}{c}{} & \multicolumn{1}{c}{0.95} & \multicolumn{1}{c}{0.95} & \multicolumn{1}{c}{0.95} & \multicolumn{1}{c}{0.94} \\
		RC    & \multicolumn{1}{c}{-0.2} & \multicolumn{1}{c}{0.3} & \multicolumn{1}{c}{-2.3} & \multicolumn{1}{c}{-1.8} & \multicolumn{1}{c}{} & \multicolumn{1}{c}{1.00} & \multicolumn{1}{c}{1.00} & \multicolumn{1}{c}{1.00} & \multicolumn{1}{c}{1.16} & \multicolumn{1}{c}{} & \multicolumn{1}{c}{0.95} & \multicolumn{1}{c}{0.94} & \multicolumn{1}{c}{0.94} & \multicolumn{1}{c}{0.93} \\
		& \multicolumn{14}{c}{DRGEE($x^-,r^+$)} \\
		ind   & \multicolumn{1}{c}{2.2} & \multicolumn{1}{c}{0.0} & \multicolumn{1}{c}{-4.7} & \multicolumn{1}{c}{-1.5} & \multicolumn{1}{c}{} & \multicolumn{1}{c}{1.00} & \multicolumn{1}{c}{1.00} & \multicolumn{1}{c}{1.00} & \multicolumn{1}{c}{1.00} & \multicolumn{1}{c}{} & \multicolumn{1}{c}{0.95} & \multicolumn{1}{c}{0.95} & \multicolumn{1}{c}{0.95} & \multicolumn{1}{c}{0.94} \\
		exch  & \multicolumn{1}{c}{0.2} & \multicolumn{1}{c}{0.6} & \multicolumn{1}{c}{-2.5} & \multicolumn{1}{c}{1.3} & \multicolumn{1}{c}{} & \multicolumn{1}{c}{0.98} & \multicolumn{1}{c}{0.99} & \multicolumn{1}{c}{1.00} & \multicolumn{1}{c}{1.17} & \multicolumn{1}{c}{} & \multicolumn{1}{c}{0.96} & \multicolumn{1}{c}{0.96} & \multicolumn{1}{c}{0.96} & \multicolumn{1}{c}{0.95} \\
		unst  & \multicolumn{1}{c}{4.7} & \multicolumn{1}{c}{-1.3} & \multicolumn{1}{c}{-11.1} & \multicolumn{1}{c}{-3.7} & \multicolumn{1}{c}{} & \multicolumn{1}{c}{1.00} & \multicolumn{1}{c}{1.01} & \multicolumn{1}{c}{1.03} & \multicolumn{1}{c}{1.18} & \multicolumn{1}{c}{} & \multicolumn{1}{c}{0.94} & \multicolumn{1}{c}{0.94} & \multicolumn{1}{c}{0.94} & \multicolumn{1}{c}{0.94} \\
		unif  & \multicolumn{1}{c}{1.2} & \multicolumn{1}{c}{0.3} & \multicolumn{1}{c}{-3.3} & \multicolumn{1}{c}{-0.1} & \multicolumn{1}{c}{} & \multicolumn{1}{c}{1.00} & \multicolumn{1}{c}{0.99} & \multicolumn{1}{c}{1.00} & \multicolumn{1}{c}{1.17} & \multicolumn{1}{c}{} & \multicolumn{1}{c}{0.96} & \multicolumn{1}{c}{0.96} & \multicolumn{1}{c}{0.97} & \multicolumn{1}{c}{0.94} \\
		cat.exch & \multicolumn{1}{c}{1.3} & \multicolumn{1}{c}{0.3} & \multicolumn{1}{c}{-3.5} & \multicolumn{1}{c}{-0.5} & \multicolumn{1}{c}{} & \multicolumn{1}{c}{1.00} & \multicolumn{1}{c}{0.99} & \multicolumn{1}{c}{1.00} & \multicolumn{1}{c}{1.18} & \multicolumn{1}{c}{} & \multicolumn{1}{c}{0.96} & \multicolumn{1}{c}{0.96} & \multicolumn{1}{c}{0.97} & \multicolumn{1}{c}{0.94} \\
		time.exch & \multicolumn{1}{c}{0.3} & \multicolumn{1}{c}{0.7} & \multicolumn{1}{c}{-1.2} & \multicolumn{1}{c}{1.3} & \multicolumn{1}{c}{} & \multicolumn{1}{c}{1.00} & \multicolumn{1}{c}{1.00} & \multicolumn{1}{c}{1.00} & \multicolumn{1}{c}{1.18} & \multicolumn{1}{c}{} & \multicolumn{1}{c}{0.96} & \multicolumn{1}{c}{0.96} & \multicolumn{1}{c}{0.97} & \multicolumn{1}{c}{0.94} \\
		RC    & \multicolumn{1}{c}{1.2} & \multicolumn{1}{c}{-0.1} & \multicolumn{1}{c}{-5.5} & \multicolumn{1}{c}{-1.6} & \multicolumn{1}{c}{} & \multicolumn{1}{c}{1.00} & \multicolumn{1}{c}{1.00} & \multicolumn{1}{c}{1.00} & \multicolumn{1}{c}{1.17} & \multicolumn{1}{c}{} & \multicolumn{1}{c}{0.95} & \multicolumn{1}{c}{0.96} & \multicolumn{1}{c}{0.96} & \multicolumn{1}{c}{0.93} \\
		& \multicolumn{14}{c}{DRGEE($x^+,r^-$)} \\
		ind   & \multicolumn{1}{c}{1.0} & \multicolumn{1}{c}{0.2} & \multicolumn{1}{c}{-2.5} & \multicolumn{1}{c}{-1.8} & \multicolumn{1}{c}{} & \multicolumn{1}{c}{1.00} & \multicolumn{1}{c}{1.00} & \multicolumn{1}{c}{1.00} & \multicolumn{1}{c}{1.00} & \multicolumn{1}{c}{} & \multicolumn{1}{c}{0.95} & \multicolumn{1}{c}{0.94} & \multicolumn{1}{c}{0.93} & \multicolumn{1}{c}{0.94} \\
		exch  & \multicolumn{1}{c}{-0.4} & \multicolumn{1}{c}{0.6} & \multicolumn{1}{c}{-1.4} & \multicolumn{1}{c}{0.5} & \multicolumn{1}{c}{} & \multicolumn{1}{c}{0.99} & \multicolumn{1}{c}{0.99} & \multicolumn{1}{c}{1.00} & \multicolumn{1}{c}{1.15} & \multicolumn{1}{c}{} & \multicolumn{1}{c}{0.94} & \multicolumn{1}{c}{0.95} & \multicolumn{1}{c}{0.96} & \multicolumn{1}{c}{0.95} \\
		unst  & \multicolumn{1}{c}{2.1} & \multicolumn{1}{c}{-0.5} & \multicolumn{1}{c}{-5.7} & \multicolumn{1}{c}{-5.6} & \multicolumn{1}{c}{} & \multicolumn{1}{c}{1.00} & \multicolumn{1}{c}{0.99} & \multicolumn{1}{c}{1.01} & \multicolumn{1}{c}{1.14} & \multicolumn{1}{c}{} & \multicolumn{1}{c}{0.93} & \multicolumn{1}{c}{0.94} & \multicolumn{1}{c}{0.94} & \multicolumn{1}{c}{0.95} \\
		unif  & \multicolumn{1}{c}{-0.2} & \multicolumn{1}{c}{0.7} & \multicolumn{1}{c}{-0.6} & \multicolumn{1}{c}{0.0} & \multicolumn{1}{c}{} & \multicolumn{1}{c}{1.00} & \multicolumn{1}{c}{0.99} & \multicolumn{1}{c}{1.00} & \multicolumn{1}{c}{1.15} & \multicolumn{1}{c}{} & \multicolumn{1}{c}{0.95} & \multicolumn{1}{c}{0.95} & \multicolumn{1}{c}{0.95} & \multicolumn{1}{c}{0.94} \\
		cat.exch & \multicolumn{1}{c}{-0.1} & \multicolumn{1}{c}{0.6} & \multicolumn{1}{c}{-0.8} & \multicolumn{1}{c}{-0.5} & \multicolumn{1}{c}{} & \multicolumn{1}{c}{1.00} & \multicolumn{1}{c}{0.99} & \multicolumn{1}{c}{1.00} & \multicolumn{1}{c}{1.15} & \multicolumn{1}{c}{} & \multicolumn{1}{c}{0.94} & \multicolumn{1}{c}{0.95} & \multicolumn{1}{c}{0.95} & \multicolumn{1}{c}{0.94} \\
		time.exch & \multicolumn{1}{c}{-0.4} & \multicolumn{1}{c}{0.8} & \multicolumn{1}{c}{0.1} & \multicolumn{1}{c}{0.5} & \multicolumn{1}{c}{} & \multicolumn{1}{c}{1.00} & \multicolumn{1}{c}{0.99} & \multicolumn{1}{c}{1.00} & \multicolumn{1}{c}{1.14} & \multicolumn{1}{c}{} & \multicolumn{1}{c}{0.94} & \multicolumn{1}{c}{0.94} & \multicolumn{1}{c}{0.95} & \multicolumn{1}{c}{0.94} \\
		RC    & \multicolumn{1}{c}{0.0} & \multicolumn{1}{c}{0.1} & \multicolumn{1}{c}{-3.3} & \multicolumn{1}{c}{-2.2} & \multicolumn{1}{c}{} & \multicolumn{1}{c}{1.00} & \multicolumn{1}{c}{1.00} & \multicolumn{1}{c}{1.00} & \multicolumn{1}{c}{1.14} & \multicolumn{1}{c}{} & \multicolumn{1}{c}{0.94} & \multicolumn{1}{c}{0.94} & \multicolumn{1}{c}{0.94} & \multicolumn{1}{c}{0.93} \\
		\hline
		\multicolumn{14}{c}{``$^+$" indicates correctly specified model and ``$^-$" indicates misspecified model omitting the $Z_1$ predictor}
	\end{tabular}%
\end{table}%

Table \ref{dois_tab2} presents simulation results for complete data in addition to correctly specified methods. In terms of bias there is no clear distinction between the two approaches, except when an unstructured matrix is chosen, which causes an increase in bias, especially for WGEE and DRGEE when only the weight model is correctly specified. For the scenario under consideration, the results suggest that the performances of both WGEE and DRGEE are slightly superior to multiple imputation, especially for estimates associated with covariate $X$.

Compared to independence structure, all other association structures presented smaller standard errors for the time-dependent covariate $Z$. Although they are very similar within each method, the largest gain in efficiency, around 23\%, is obtained for the MIGEE method, followed by WGEE with 19\%. For DRGEE estimators it ranged from 14\% to 18\%.

All correctly specified methods showed coverage rates close to the nominal levels for all association structures. In terms of empirical bias, relative efficiency and empirical coverage there was no clear distinction between the correlation and local odds parametrizations for $n=300$ subjects. For the scenario under consideration, an exchangeable correlation structure or an uniform local odds structure resulted in good marginal mean estimates.

Simulation results for $n=50, 150$ and $600$ are presented in the Appendix. With sample sizes $n=50$ or $n=150$ subjects, simulation results suggest that the local odds parametrization outperforms the correlation coefficient parametrization in terms of bias and convergence issues.

\section{Data Analysis: Functional Classification in Rheumatic Mitral Stenosis}\label{dois_realdata}

A cohort of 164 patients with rheumatic mitral stenosis who were referred for treatment at Hospital das Clinicas of the Federal University of Minas Gerais, Brazil, was selected for a mitral valve invervention. Patients were included before intervention and then followed up in the outpatient clinic every $4$ months according to their clinical status. The first three measurements were available for analysis.  

Mitral stenosis is a narrowing of the mitral valve in the heart caused by rheumatic disease, which restricts the flow of blood through the valve. The main clinical manifestation of this disease is shortness of breath, classified in four categories based on how much the patients are limited during physical activity. The response of interest is The New York Heart Association (NYHA) Functional Classification, that provides a simple way of classifying the extent of shortness of breath. Patients with no symptoms and no limitation in ordinary physical activity were classified in class I; slight limitation of physical activity in class II; marked limitation of physical activity in class III; and patients with severe limitations resulting in inability to carry on any physical activity without discomfort in class IV. Only one patient were classified into class IV in the followup evaluation and hence class IV was combined with class III in the analysis. Thus, the ordinal response was defined as ($1$: if class I, $2$: if class II, $3$ if class III or class IV).

Percutaneous mitral valvuloplasty (PMV) is an effective treatment for stretching the stenosed mitral valve. This procedure is carried out by inserting a catheter with a balloon at its tip to open the narrowed mitral valve. This procedure causes improvement of the functional class in the majority of the patients. A number of patient's characteristics were measured at baseline, such as atrial compliance (Cn: defined as $1$, if $\leq 4$, and $0$ otherwise), cardiac rhythm, morphological features of the mitral valve expressed as an echocardiographic score, mitral valve area, pressure transmitral gradients, and pulmonary artery pressure. Some variables measured after the procedure include the success of the procedure to open the mitral valve without complications, long-term event-free survival, mitral valve area, pressure transmitral gradients, and pulmonary artery pressure at the follow-up appointment.

This study was characterized by an arbitrary pattern of missing data. The response were fully observed for $125$ patients, $29$ had only the first measurement, $1$ patient had only the first data collected, and for $9$ patients the second occasion was missing. There was no missing data in response at baseline. Some collected variables, such as success of the procedure and long term events were responsible for the missingness at followup. A baseline covariate of particular interest, atrial compliance, were missing for $54 (32.9\%)$ of the patients. Among the reasons for not observing such predictor it can be included morphological characteristics of the mitral valve and valve calcification. Therefore a MAR mechanism seems to be a reasonable assumption for this data set. 

Now, the models used for analysis are described. For the ordinal response it was used the following proportional odds model
\begin{equation}\label{mod.resp}
	logit \ Pr(NYHA_{itj}\leq j|\boldsymbol{u_{it}})=\beta_{0j}+\boldsymbol{u}_{it}^T\boldsymbol{\beta}, \ \ \ j=1,2, \ \ t=1,2,3,
\end{equation} 
where $\boldsymbol{u}_{it}$ is the covariate vector at time $t$, and it is formed by time, atrial compliance, echocardiographic score, and success of the procedure. 

When using WGEE or DRGGE it is necessary to correctly model the missingness mechanism in order to obtain consistent estimates of $\boldsymbol{\beta}$. For the missing data process, $R_{it}^y$ was defined as the indicator of observing the response $NYHA_{it}$ and $R_i^x$ was defined as the indicator of observing the baseline atrial compliance. The probability of observing the response was modeled as
\begin{equation}\label{mod.perda.y}
	logit \ Pr(R_{it}^y=1)=\psi_0^y+\boldsymbol{w}_{it}^{yT}\boldsymbol{\psi}^y, \ \ t=2,3,
\end{equation}
where the vector $\boldsymbol{w}_{it}^y$ included success of the procedure, long term events, and the previous indicator of missing data. The model for $R_i^x$ was specified as
\begin{equation}\label{mod.perda.x}
	logit \ Pr(R_i^x=1)=\psi_0^x+\boldsymbol{w}_{i}^{xT}\boldsymbol{\psi}^x,
\end{equation}
where the vector $\boldsymbol{w}_{i}^x$ included the baseline response, echocardiographic score, valve calcification and ECG rhythm.

For the covariate, the following model was built
\begin{equation}\label{mod.cov.x}
	logit \ Pr(Cn=1)=\boldsymbol{v}_i^T\boldsymbol{\gamma},
\end{equation}
where $\boldsymbol{v}_i^T$ included the baseline right ventricular systolic pressure, mean gradient and mitral valve area. A imputation model for NYHA was also specified that included the covariates in (\ref{mod.resp}) in addition to the baseline right ventricular systolic pressure and the response history. 

Here, four methods were used to analyze the data. Results are shown in Table \ref{dois_tab3}.  The first method is the usual GEE method using the available data; the second is the weighted method (WGEE) using models (\ref{mod.perda.y}) and (\ref{mod.perda.x}) for the weights; the third is the multiple imputation by chained equation (MIGEE); and the fourth, labeled DRGEE, is the doubly robust method using (\ref{mod.perda.y}) and (\ref{mod.perda.x}) for the missing data process and (\ref{mod.cov.x}) for the covariate models. In order to account for the dependence structure of the repeated measures the correlation coefficient and the local odds ratio parametrization were both initially applied with the same association structures presented in the simulation. Convergence issues were observed for unstructured correlation matrix. Motivated by simulation results that suggested better performance of the local odds ratio structures, results are shown only for the independence, uniform and category exchangeability local odds ratio structures. The independence structure estimates no parameters, while the uniform represents the dependence structure by a single parameter. The category exchangeability assumes constant odds ratio among the levels of the response, but different odds ratio between time pairs. Thus, the association is explained by three parameters in this structure. Similar results were obtained for the other dependence structures they are shown in Appendix.

\begin{table}[htbp]
	\centering
	\caption{Results for Rheumatic Mitral Stenosis study under independence, uniform and category exchangeability association structures}\label{dois_tab3}
	\begin{tabular}{rcccrcccrccc}
		\hline
		& \multicolumn{3}{c}{Independence} &       & \multicolumn{3}{c}{Uniform} &       & \multicolumn{3}{c}{Cat. Exchangeability} \\
		\cline{2-4}\cline{6-8}\cline{10-12}    Parameter & Est.  & SE    & p     &       & Est.  & SE    & p     &       & Est.  & SE    & p \\
		\cline{1-4}\cline{6-8}\cline{10-12}    \textbf{} & \multicolumn{11}{c}{Available} \\
		$\beta_{01}$ & -0.800 & 0.765 & 0.295 &       & -0.838 & 0.735 & 0.254 &       & -0.768 & 0.751 & 0.307 \\
		$\beta_{02}$  & 0.997 & 0.745 & 0.181 &       & 0.954 & 0.726 & 0.189 &       & 1,037 & 0.746 & 0.164 \\
		Success & 0.733 & 0.441 & 0.096 &       & 0.662 & 0.391 & 0.090 &       & 0.692 & 0.400 & 0.083 \\
		Total Score & -0.147 & 0.104 & 0.158 &       & -0.142 & 0.101 & 0.161 &       & -0.156 & 0.103 & 0.130 \\
		Cn    & 0.533 & 0.328 & 0.104 &       & 0.498 & 0.315 & 0.113 &       & 0.499 & 0.318 & 0.116 \\
		time=2 & 1,003 & 0.413 & 0.015 &       & 1,084 & 0.389 & 0.005 &       & 1,126 & 0.390 & 0.004 \\
		time=3 & 1,216 & 0.420 & 0.004 &       & 1,214 & 0.390 & 0.002 &       & 1,198 & 0.394 & 0.002 \\
		& \multicolumn{11}{c}{WGEE} \\
		$\beta_{01}$  & -0.852 & 0.786 & 0.278 &       & -0.883 & 0.763 & 0.247 &       & -0.792 & 0.776 & 0.307 \\
		$\beta_{02}$  & 0.895 & 0.768 & 0.244 &       & 0.882 & 0.748 & 0.238 &       & 0.974 & 0.767 & 0.204 \\
		Success & 0.750 & 0.449 & 0.094 &       & 0.732 & 0.405 & 0.071 &       & 0.749 & 0.416 & 0.072 \\
		Total Score & -0.139 & 0.107 & 0.192 &       & -0.136 & 0.104 & 0.190 &       & -0.153 & 0.106 & 0.148 \\
		Cn    & 0.524 & 0.336 & 0.118 &       & 0.464 & 0.325 & 0.153 &       & 0.472 & 0.327 & 0.148 \\
		time=2 & 1,030 & 0.413 & 0.013 &       & 1,039 & 0.401 & 0.010 &       & 1,085 & 0.406 & 0.007 \\
		time=3 & 1,269 & 0.419 & 0.002 &       & 1,352 & 0.404 & 0.001 &       & 1,331 & 0.410 & 0.001 \\
		& \multicolumn{11}{c}{MIGEE} \\
		$\beta_{01}$  & -0.856 & 0.582 & 0.141 &       & -0.805 & 0.580 & 0.165 &       & -0.788 & 0.572 & 0.169 \\
		$\beta_{02}$  & 0.988 & 0.582 & 0.090 &       & 1,019 & 0.574 & 0.076 &       & 1,035 & 0.572 & 0.070 \\
		Success & 0.938 & 0.367 & 0.010 &       & 0.922 & 0.333 & 0.006 &       & 0.980 & 0.338 & 0.004 \\
		Total Score & -0.141 & 0.080 & 0.077 &       & -0.140 & 0.079 & 0.075 &       & -0.147 & 0.078 & 0.060 \\
		Cn    & 0.425 & 0.305 & 0.163 &       & 0.307 & 0.261 & 0.239 &       & 0.343 & 0.267 & 0.199 \\
		time=2 & 0.995 & 0.334 & 0.003 &       & 1,106 & 0.308 & 0.000 &       & 1,127 & 0.330 & 0.001 \\
		time=3 & 0.969 & 0.338 & 0.004 &       & 1,068 & 0.304 & 0.000 &       & 1,079 & 0.333 & 0.001 \\
		& \multicolumn{11}{c}{DRGEE} \\
		$\beta_{01}$  & -0.793 & 0.730 & 0.277 &       & -0.830 & 0.711 & 0.243 &       & -0.726 & 0.711 & 0.308 \\
		$\beta_{02}$  & 0.952 & 0.722 & 0.187 &       & 0.934 & 0.707 & 0.186 &       & 1.039  & 0.712 & 0.145 \\
		Success & 0.825 & 0.422 & 0.050 &       & 0.806 & 0.382 & 0.035 &       & 0.841 & 0.388 & 0.030 \\
		Total Score & -0.150 & 0.099 & 0.131 &       & -0.147 & 0.097 & 0.132 &       & -0.165 & 0.098 & 0.092 \\
		Cn    & 0.571 & 0.341 & 0.094 &       & 0.520 & 0.329 & 0.113 &       & 0.523 & 0.333 & 0.116 \\
		time=2 & 1.013 & 0.499 & 0.042 &       & 1.005 & 0.512 & 0.050 &       & 1.020  & 0.525 & 0.052 \\
		time=3 & 1.135 & 0.476 & 0.017 &       & 1.185 & 0.473 & 0.012 &       & 1.145  & 0.491 & 0.020\\
		\hline
	\end{tabular}%
	\label{tab:addlabel}%
\end{table}%

The significance of time effect indicates the effectiveness of the intervention and improvement of functional class over time. Similar coefficients for the two followup occasions suggest the major change in functional class occurs right after the valvuloplasty intervention. Total score was non significant for all the four methods, although a marginally significance ($p=0.060$) was noted for MIGEE when the category exchangeability odds ratio structure was adopted, as an indication that higher scores may be related to cardiac insufficiency. All methods provide the same conclusion for effects of the missing covariate Cn. It can be noticed that estimates for success of the procedure effect goes from non significant in the standard GEE to significant for all methods, except WGEE, regardless the adopted association structure. It is interesting to note that the estimated effect are increased for all methods, especially for the multiple imputation procedure. So, for example, for the uniform structure, the estimated odds of response in class I for subjects with success of the procedure compared to patients with suboptimal results were $e^{0.806}=2.24$ for DRGEE and $e^{0.922}=2.51$ for MIGEE. Regarding the association structures it can be noticed that the uniform and category exchangeability choices presented some gain in efficiency, specially for the success of procedure effect. Similar conclusions were reached for the other association structures.

\section{Discussion}\label{dois_discuss}

In this paper it was considered a doubly robust estimator for analysis of longitudinal data when missingness can occur in a baseline covariate or intermittently in the ordinal response. The main objective was to compare the performance of the proposed method in terms of bias and efficiency for two different approaches to modeling the covariance matrix of the longitudinal outcome, namely, the correlation coefficient proposed by \cite{lipsitz1994} and local odds ratio proposed by \cite{touloumis2013}. Although a complete data comparison between these two distinct approaches had already appeared elsewhere \citep{nooraee2014} such a comparison with MAR missing data was still in need of investigation.

The covariate design plays an important role in the efficiency of GEE estimators. The working independence structure is expected to be efficient for time-stationary covariates. This is no longer true for time-varying covariates and/or missing data  \citep{lipsitz1994}. Simulation results agreed with the literature, that is, the gain in efficiency by adopting more complicated association structures was noticed for the time-varying covariate and it somewhat varied through the different methods. For complete data, the standard error for the independence was, on average, about 27\% larger compared to an uniform structure that estimates a single association parameter. The gain in efficiency reached 23\% for MIGEE, 19\% for WGEE and 18\% for DRGEE. 

As \cite{liang1986} pointed out when the assumed correlation is the true correlation, the missing completely at random assumption can be unnecessary. However, this is not true when missingness occurs in a covariate that is MAR given the response. In this case even likelihood methods are biased \citep{carpenter2013} and the missingness mechanism must be modeled in order to obtain valid estimates for the regression parameters.

The dependence structures compared here differ in terms of number of parameters and restrictions imposed on the correlation/association between levels of ordinal response at different time pairs. Although the same definition can be used for an independence, exchangeable and unstructured association matrix, this does not imply that identical associations are fit \citep{nooraee2014}. Under the independence working assumption all off-diagonal blocks of the covariance matrix are constant and equal to zero. Exchangeability over time indicates that the association between $Y_{itj}$ and $Y_{it'j'}$ is independent of time, but it depends on the levels $j$ and $j'$. For unstructured associations there are no restrictions implied. With moderate to large number of subjects (at least $300$ subjects) the simulation results did not suggest a very clear distinction in terms of bias, relative efficiency and empirical coverage between the correlation or local odds ratio parametrizations, although some bias was observed for unstructured matrices. On the other side, for small sample sizes ($n=50$ or $150$) simulation results did suggest that the local odds ratio outperforms the correlation coefficient parametrization in terms of bias and convergence issues. It seems that local odds structures works fine for small samples sizes and the correlation coefficient needs relatively more subjects to achieve the same reduction in bias.

It is noticeable two important differences between the local odds ratio and the correlation coefficient parametrizations. Firstly, the local odds ratios does not depend on the marginal specification, that is, $\boldsymbol{\beta}$ and $\boldsymbol{\alpha}$ are variation independent \citep{touloumis2013}. Unlike correlations, the local odds estimates for the association structure does not depend on the values of the observed covariates and thus these estimates do not need to be obtained at each step of the modified Fisher Scoring algorithm. As a consequence, the iterative procedure converges faster. Secondly, the correlation estimates are obtained through a moment based estimator, thus some work was necessary to ensure their validity in the presence of data that is MAR. The estimates of local odds are based on maximum likelihood methods, assuming independent Poisson sampling for the observed counts in the marginalized contingency tables. As it considers only the observed responses the resulting estimates are valid under a MAR mechanism. 

In some situations (small samples, complex patterns of missing data), it is possible that the algorithm will not converge because the estimated correlation matrix is not guaranteed to be positive definite \citep{lipsitz1994}. For small sample sizes there may not be enough data to estimate both the regression parameters and a correlation matrix that is highly unstructured. Simulation results for unstructured correlation matrices and small sample sizes presented very low convergence rates (in addition to increased bias) and thus an exchangeable correlation matrix is preferred. There were no serious convergence issues with the local odds ratio parametrization, except for an unstructured matrix applied to $n=50$.

Although the latent vectors have been generated from an exchangeable correlation, the correlation coefficient between binary variables $Y_{itj}$ and $Y_{it'j'}$ also depends on the linear predictor in times $t$ and $t'$ \citep{nooraee2014}. That is, the correlation between the latent vectors was exchangeable, but the correlation between the binary variables was not exchangeable due to the dependent mean in the marginal model. Nevertheless, it is expected the GEE method remain valid even assuming a misspecified correlation matrix provided the marginal mean is correctly specified \citep{molenberghs2010semi}. 

Overall, the doubly robust method performed well for all working association structures under comparison except when applied to a very small number of subjects (i.e., $n=50$) and an unstructured matrix was adopted. The coverage probabilities were relatively close to the nominal level of 95\%.

The doubly robust estimator considered here is restricted to missingness in the response and a baseline covariate. A natural extension is to consider an intermittently missing time-varying covariate and/or allow multiple missing covariates. In the proposed approach the marginal means were modeled by cumulative logits. This implies a  proportional odds model that in some cases may not be valid. Another possible extension of the proposed model is, therefore, to allow non-proportional odds for a subset of the explanatory variables \citep{peterson1990}.

\addcontentsline{toc}{section}{References}
\bibliographystyle{authordate1}
\bibliography{biblio}

\newpage
\appendix
\section{Appendix}
\subsection{Asymptotic Variance}
To state the asymptotic properties of $\boldsymbol{\hat{\beta}}$, let\\
$\boldsymbol{S}_{1i}(\boldsymbol{\beta},\boldsymbol{\psi},\boldsymbol{\gamma})$ be the individual's contribution to the estimating equations for $\boldsymbol{\beta}$,\\
$\boldsymbol{S}_{2i}(\boldsymbol{\psi})$ be the individual's contribution to the  estimating equations for  $\boldsymbol{\psi}$, and\\
$\boldsymbol{S}_{3i}(\boldsymbol{\gamma})$ be the individual's contribution to the estimating equations for  $\boldsymbol{\gamma}$.

Define 
$\boldsymbol{\Gamma}(\boldsymbol{\beta},\boldsymbol{\psi},\boldsymbol{\gamma})=
E\left\{ \partial \boldsymbol{S}_{1i}(\boldsymbol{\beta},\boldsymbol{\psi},\boldsymbol{\gamma})/ \partial \boldsymbol{\beta}^T   \right\}$,
$\boldsymbol{I}_{12}(\boldsymbol{\beta},\boldsymbol{\psi},\boldsymbol{\gamma})=E\left\{ \partial \boldsymbol{S}_{1i}(\boldsymbol{\beta},\boldsymbol{\psi},\boldsymbol{\gamma})/ \partial \boldsymbol{\psi}^T   \right\}$,
$\boldsymbol{I}_{13}(\boldsymbol{\beta},\boldsymbol{\psi},\boldsymbol{\gamma})=E\left\{ \partial \boldsymbol{S}_{1i}(\boldsymbol{\beta},\boldsymbol{\psi},\boldsymbol{\gamma})/ \partial \boldsymbol{\gamma}^T  \right\}$, $\boldsymbol{I}_{2}(\boldsymbol{\psi})=E\left\{ \partial \boldsymbol{S}_{2i}(\boldsymbol{\psi})/ \partial \boldsymbol{\psi}^T   \right\}$,
$\boldsymbol{I}_{3}(\boldsymbol{\gamma})=E\left\{ \partial \boldsymbol{S}_{3i}(\boldsymbol{\gamma})/ \partial \boldsymbol{\gamma}^T   \right\}$, and
$\boldsymbol{Q}_i(\boldsymbol{\beta},\boldsymbol{\psi},\boldsymbol{\gamma})=\boldsymbol{S}_{1i}(\boldsymbol{\beta},\boldsymbol{\psi},\boldsymbol{\gamma})-
\boldsymbol{I}_{12}(\boldsymbol{\beta},\boldsymbol{\psi},\boldsymbol{\gamma})\boldsymbol{I}_{2}^{-1}(\boldsymbol{\psi})\boldsymbol{S}_{2i}(\boldsymbol{\psi})-
\boldsymbol{I}_{13}(\boldsymbol{\beta},\boldsymbol{\psi},\boldsymbol{\gamma})\boldsymbol{I}_{3}^{-1}(\boldsymbol{\gamma})\boldsymbol{S}_{3i}(\boldsymbol{\gamma})$.

\begin{theo}
	If either the missing data model or the covariate model is correctly specified, then
	\begin{equation}\label{vardr}
	n^{1/2}(\boldsymbol{\hat{\beta}}-\boldsymbol{\beta_0})\longrightarrow N(\boldsymbol{0},\boldsymbol{\Gamma}^{-1}(\boldsymbol{\beta}_0,\boldsymbol{\psi}_0,\boldsymbol{\gamma}_0)\boldsymbol{\Sigma}
	\left\{\boldsymbol{\Gamma}^{-1}(\boldsymbol{\beta}_0,\boldsymbol{\psi}_0,\boldsymbol{\gamma}_0)\right\}^T),
	\end{equation}
	where $\boldsymbol{\beta}_0$ is the true value of $\boldsymbol{\beta}$, $\boldsymbol{\psi}_0$ and $\boldsymbol{\gamma}_0$ are the probability limits of $\boldsymbol{\hat{\psi}}$ and $\boldsymbol{\hat{\gamma}}$, and \\ $\boldsymbol{\Sigma}=E\left\{\boldsymbol{Q}_i(\boldsymbol{\beta}_0,\boldsymbol{\psi}_0,\boldsymbol{\gamma}_0)\boldsymbol{Q}_i^T(\boldsymbol{\beta}_0,\boldsymbol{\psi}_0,\boldsymbol{\gamma}_0)    \right\}$.
\end{theo}

Inferences for $\boldsymbol{\beta}$ follows by replacing the unknown quantities in (\ref{vardr}) by its consistent estimators. We make use of ``generalized information equality'' \citep{pierce1982} that \\
$E\left\{ \partial \boldsymbol{S}_{1i}(\boldsymbol{\beta},\boldsymbol{\psi},\boldsymbol{\gamma})/ \partial \boldsymbol{\psi}^T  \right\}=-
E\left\{ \boldsymbol{S}_{1i}(\boldsymbol{\beta},\boldsymbol{\psi},\boldsymbol{\gamma})\boldsymbol{S}_{2i}^T(\boldsymbol{\psi})\right\}$, and \\
$E\left\{ \partial \boldsymbol{S}_{1i}(\boldsymbol{\beta},\boldsymbol{\psi},\boldsymbol{\gamma})/ \partial \boldsymbol{\gamma}^T  \right\}=-
E\left\{ \boldsymbol{S}_{1i}(\boldsymbol{\beta},\boldsymbol{\psi},\boldsymbol{\gamma})\boldsymbol{S}_{3i}^T(\boldsymbol{\gamma})\right\}$. Similarly \citep{robins1995},\\
$E\left\{ \partial \boldsymbol{S}_{2i}(\boldsymbol{\psi})/ \partial \boldsymbol{\psi}^T  \right\}=-
Var\left\{\boldsymbol{S}_{2i}(\boldsymbol{\psi})  \right\}$, and
$E\left\{ \partial \boldsymbol{S}_{3i}(\boldsymbol{\gamma})/ \partial \boldsymbol{\gamma}^T  \right\}=-
Var\left\{\boldsymbol{S}_{3i}(\boldsymbol{\gamma})  \right\}$.

The matrix $\boldsymbol{\Gamma}$ is replaced by $\boldsymbol{\hat{\Gamma}}=n^{-1}\sum_{i=1}^n\left\{ \partial \boldsymbol{S}_{1i}(\boldsymbol{\hat{\theta}})/\partial \boldsymbol{\beta}^T  \right\}$, and 
$\boldsymbol{\Sigma}$ by $\boldsymbol{\hat{\Sigma}}=n^{-1} \sum_{i=1}^n \left\{\boldsymbol{\hat{Q}}_i\boldsymbol{\hat{Q}}_i^T  \right\}$,
$\boldsymbol{\hat{Q}}_i=\boldsymbol{S}_{1i}(\boldsymbol{\hat{\theta}})-\boldsymbol{\hat{I}}_{12}(\boldsymbol{\hat{\theta}})\boldsymbol{\hat{I}}_2^{-1}(\boldsymbol{\hat{\psi}})\boldsymbol{S}_{2i}(\boldsymbol{\hat{\psi}})-\boldsymbol{\hat{I}}_{13}(\boldsymbol{\hat{\theta}})\boldsymbol{\hat{I}}_3^{-1}(\boldsymbol{\hat{\gamma}})\boldsymbol{S}_{3i}(\boldsymbol{\hat{\gamma}})$,
$\boldsymbol{\hat{I}}_{12}(\boldsymbol{\hat{\theta}})=n^{-1}\sum_{i=1}^n\left\{ \partial \boldsymbol{S}_{1i}(\boldsymbol{\hat{\theta}})/ \partial \boldsymbol{\psi}^T   \right\}$,
$\boldsymbol{\hat{I}}_{13}(\boldsymbol{\hat{\theta}})=n^{-1}\sum_{i=1}^n\left\{ \partial \boldsymbol{S}_{1i}(\boldsymbol{\hat{\theta}})/ \partial \boldsymbol{\gamma}^T \right\}$,
$\boldsymbol{\hat{I}}_{2}(\boldsymbol{\hat{\psi}})=n^{-1}\sum_{i=1}^n\left\{ \partial \boldsymbol{S}_{2i}(\boldsymbol{\hat{\psi}})/ \partial \boldsymbol{\psi}^T \right\}$, \\
$\boldsymbol{\hat{I}}_{3}(\boldsymbol{\hat{\gamma}})=n^{-1}\sum_{i=1}^n\left\{ \partial \boldsymbol{S}_{3i}(\boldsymbol{\hat{\gamma}})/ \partial \boldsymbol{\gamma}^T \right\}$.

The proof is similar to \cite{chen2011} and is omitted here.

\newpage
\subsection{Additional Simulation Results}

\begin{table}[htbp]
	\centering
	\scriptsize
	\caption{Evaluation criteria for misspecified models. Results for $n=50$ and $S=1000$ simulations.}\label{dois_n50_wro}
\begin{tabular}{rrrrrrrrrrrrrrr}
		\hline
		& \multicolumn{4}{c}{Relative Bias} &       & \multicolumn{4}{c}{Relative Efficiency} &       & \multicolumn{4}{c}{Empirical Coverage} \\
		\cline{2-5}\cline{7-10}\cline{12-15}    Structure & $\beta_{01}$ & $\beta_{02}$ & $X$     & $Z$     &       & $\beta_{01}$ & $\beta_{02}$ & $X$     & $Z$     &       & $\beta_{01}$ & $\beta_{02}$ & $X$     & $Z$ \\
		\hline
		& \multicolumn{14}{c}{Available} \\
		ind   & \multicolumn{1}{c}{101.9} & \multicolumn{1}{c}{-25.2} & \multicolumn{1}{c}{-10.0} & \multicolumn{1}{c}{48.4} & \multicolumn{1}{c}{} & \multicolumn{1}{c}{1.00} & \multicolumn{1}{c}{1.00} & \multicolumn{1}{c}{1.00} & \multicolumn{1}{c}{1.00} & \multicolumn{1}{c}{} & \multicolumn{1}{c}{0.85} & \multicolumn{1}{c}{0.85} & \multicolumn{1}{c}{1.00} & \multicolumn{1}{c}{0.95} \\
		exch  & \multicolumn{1}{c}{96.2} & \multicolumn{1}{c}{-22.6} & \multicolumn{1}{c}{-36.0} & \multicolumn{1}{c}{19.5} & \multicolumn{1}{c}{} & \multicolumn{1}{c}{1.02} & \multicolumn{1}{c}{1.03} & \multicolumn{1}{c}{1.03} & \multicolumn{1}{c}{1.15} & \multicolumn{1}{c}{} & \multicolumn{1}{c}{0.87} & \multicolumn{1}{c}{0.88} & \multicolumn{1}{c}{1.00} & \multicolumn{1}{c}{0.95} \\
		unst  & \multicolumn{1}{c}{91.0} & \multicolumn{1}{c}{-21.8} & \multicolumn{1}{c}{-44.1} & \multicolumn{1}{c}{-1.3} & \multicolumn{1}{c}{} & \multicolumn{1}{c}{1.08} & \multicolumn{1}{c}{1.10} & \multicolumn{1}{c}{1.10} & \multicolumn{1}{c}{1.17} & \multicolumn{1}{c}{} & \multicolumn{1}{c}{0.86} & \multicolumn{1}{c}{0.90} & \multicolumn{1}{c}{1.00} & \multicolumn{1}{c}{0.96} \\
		unif  & \multicolumn{1}{c}{80.2} & \multicolumn{1}{c}{-18.6} & \multicolumn{1}{c}{-14.8} & \multicolumn{1}{c}{26.5} & \multicolumn{1}{c}{} & \multicolumn{1}{c}{1.01} & \multicolumn{1}{c}{0.99} & \multicolumn{1}{c}{1.00} & \multicolumn{1}{c}{1.19} & \multicolumn{1}{c}{} & \multicolumn{1}{c}{0.91} & \multicolumn{1}{c}{0.90} & \multicolumn{1}{c}{1.00} & \multicolumn{1}{c}{0.94} \\
\end{tabular}%
\end{table}%

\begin{table}[htbp]
	\centering
	\scriptsize
	\caption{Evaluation criteria for correctly specified models. Results for $n=50$ and $S=1000$ simulations.}\label{dois_n50_cor}
	\begin{tabular}{rrrrrrrrrrrrrrr}
		\hline
		& \multicolumn{4}{c}{Relative Bias} &       & \multicolumn{4}{c}{Relative Efficiency} &       & \multicolumn{4}{c}{Empirical Coverage} \\
		\cline{2-5}\cline{7-10}\cline{12-15}    Structure & $\beta_{01}$ & $\beta_{02}$ & $X$     & $Z$     &       & $\beta_{01}$ & $\beta_{02}$ & $X$     & $Z$     &       & $\beta_{01}$ & $\beta_{02}$ & $X$     & $Z$ \\
		\hline
		& \multicolumn{14}{c}{Complete} \\
		ind   & \multicolumn{1}{c}{3.4} & \multicolumn{1}{c}{4.4} & \multicolumn{1}{c}{3.3} & \multicolumn{1}{c}{3.5} & \multicolumn{1}{c}{} & \multicolumn{1}{c}{1.00} & \multicolumn{1}{c}{1.00} & \multicolumn{1}{c}{1.00} & \multicolumn{1}{c}{1.00} & \multicolumn{1}{c}{} & \multicolumn{1}{c}{0.95} & \multicolumn{1}{c}{0.94} & \multicolumn{1}{c}{0.99} & \multicolumn{1}{c}{0.96} \\
		exch  & \multicolumn{1}{c}{13.6} & \multicolumn{1}{c}{1.3} & \multicolumn{1}{c}{-12.4} & \multicolumn{1}{c}{-8.0} & \multicolumn{1}{c}{} & \multicolumn{1}{c}{0.99} & \multicolumn{1}{c}{1.01} & \multicolumn{1}{c}{1.01} & \multicolumn{1}{c}{1.23} & \multicolumn{1}{c}{} & \multicolumn{1}{c}{0.94} & \multicolumn{1}{c}{0.96} & \multicolumn{1}{c}{1.00} & \multicolumn{1}{c}{0.94} \\
		unst  & \multicolumn{1}{c}{17.0} & \multicolumn{1}{c}{2.3} & \multicolumn{1}{c}{-25.5} & \multicolumn{1}{c}{-18.9} & \multicolumn{1}{c}{} & \multicolumn{1}{c}{1.03} & \multicolumn{1}{c}{1.06} & \multicolumn{1}{c}{1.05} & \multicolumn{1}{c}{1.21} & \multicolumn{1}{c}{} & \multicolumn{1}{c}{0.95} & \multicolumn{1}{c}{0.96} & \multicolumn{1}{c}{0.98} & \multicolumn{1}{c}{0.94} \\
		unif  & \multicolumn{1}{c}{0.1} & \multicolumn{1}{c}{4.6} & \multicolumn{1}{c}{4.0} & \multicolumn{1}{c}{2.8} & \multicolumn{1}{c}{} & \multicolumn{1}{c}{1.00} & \multicolumn{1}{c}{1.00} & \multicolumn{1}{c}{1.00} & \multicolumn{1}{c}{1.25} & \multicolumn{1}{c}{} & \multicolumn{1}{c}{0.95} & \multicolumn{1}{c}{0.93} & \multicolumn{1}{c}{1.00} & \multicolumn{1}{c}{0.95} \\
\end{tabular}%
\end{table}%

\begin{table}[htbp]
	\centering
	\scriptsize
	\caption{Evaluation criteria for misspecified models. Results for $n=150$ and $S=1000$ simulations.}\label{dois_n150_wro}
	\begin{tabular}{rrrrrrrrrrrrrrr}
		\hline
		& \multicolumn{4}{c}{Relative Bias} &       & \multicolumn{4}{c}{Relative Efficiency} &       & \multicolumn{4}{c}{Empirical Coverage} \\
		\cline{2-5}\cline{7-10}\cline{12-15}    Structure & $\beta_{01}$ & $\beta_{02}$ & $X$     & $Z$     &       & $\beta_{01}$ & $\beta_{02}$ & $X$     & $Z$     &       & $\beta_{01}$ & $\beta_{02}$ & $X$     & $Z$ \\
		\hline
		& \multicolumn{14}{c}{Available} \\
		ind   & \multicolumn{1}{c}{117.0} & \multicolumn{1}{c}{-29.9} & \multicolumn{1}{c}{-20.9} & \multicolumn{1}{c}{60.5} & \multicolumn{1}{c}{} & \multicolumn{1}{c}{1.00} & \multicolumn{1}{c}{1.00} & \multicolumn{1}{c}{1.00} & \multicolumn{1}{c}{1.00} & \multicolumn{1}{c}{} & \multicolumn{1}{c}{0.46} & \multicolumn{1}{c}{0.66} & \multicolumn{1}{c}{0.94} & \multicolumn{1}{c}{0.84} \\
		exch  & \multicolumn{1}{c}{96.4} & \multicolumn{1}{c}{-25.5} & \multicolumn{1}{c}{-32.2} & \multicolumn{1}{c}{30.2} & \multicolumn{1}{c}{} & \multicolumn{1}{c}{1.02} & \multicolumn{1}{c}{1.01} & \multicolumn{1}{c}{1.02} & \multicolumn{1}{c}{1.20} & \multicolumn{1}{c}{} & \multicolumn{1}{c}{0.58} & \multicolumn{1}{c}{0.7} & \multicolumn{1}{c}{0.92} & \multicolumn{1}{c}{0.92} \\
		unst  & \multicolumn{1}{c}{97.4} & \multicolumn{1}{c}{-26.3} & \multicolumn{1}{c}{-40.0} & \multicolumn{1}{c}{17.4} & \multicolumn{1}{c}{} & \multicolumn{1}{c}{1.04} & \multicolumn{1}{c}{1.03} & \multicolumn{1}{c}{1.04} & \multicolumn{1}{c}{1.21} & \multicolumn{1}{c}{} & \multicolumn{1}{c}{0.59} & \multicolumn{1}{c}{0.71} & \multicolumn{1}{c}{0.94} & \multicolumn{1}{c}{0.94} \\
		unif  & \multicolumn{1}{c}{92.5} & \multicolumn{1}{c}{-22.6} & \multicolumn{1}{c}{-25.9} & \multicolumn{1}{c}{26.8} & \multicolumn{1}{c}{} & \multicolumn{1}{c}{1.02} & \multicolumn{1}{c}{1.00} & \multicolumn{1}{c}{1.01} & \multicolumn{1}{c}{1.23} & \multicolumn{1}{c}{} & \multicolumn{1}{c}{0.63} & \multicolumn{1}{c}{0.78} & \multicolumn{1}{c}{0.93} & \multicolumn{1}{c}{0.91} \\
\end{tabular}%
\end{table}%

\begin{table}[htbp]
	\centering
	\scriptsize
	\caption{Evaluation criteria for correctly specified models. Results for $n=150$ and $S=1000$ simulations.}\label{dois_n150_cor}
	\begin{tabular}{rrrrrrrrrrrrrrr}
		\hline
		& \multicolumn{4}{c}{Relative Bias} &       & \multicolumn{4}{c}{Relative Efficiency} &       & \multicolumn{4}{c}{Empirical Coverage} \\
		\cline{2-5}\cline{7-10}\cline{12-15}    Structure & $\beta_{01}$ & $\beta_{02}$ & $X$     & $Z$     &       & $\beta_{01}$ & $\beta_{02}$ & $X$     & $Z$     &       & $\beta_{01}$ & $\beta_{02}$ & $X$     & $Z$ \\
		\hline
		& \multicolumn{14}{c}{Complete} \\
		ind   & \multicolumn{1}{c}{0.2} & \multicolumn{1}{c}{1.9} & \multicolumn{1}{c}{2.9} & \multicolumn{1}{c}{1.0} & \multicolumn{1}{c}{} & \multicolumn{1}{c}{1.00} & \multicolumn{1}{c}{1.00} & \multicolumn{1}{c}{1.00} & \multicolumn{1}{c}{1.00} & \multicolumn{1}{c}{} & \multicolumn{1}{c}{0.94} & \multicolumn{1}{c}{0.94} & \multicolumn{1}{c}{0.95} & \multicolumn{1}{c}{0.95} \\
		exch  & \multicolumn{1}{c}{0.9} & \multicolumn{1}{c}{0.8} & \multicolumn{1}{c}{-3.8} & \multicolumn{1}{c}{0.7} & \multicolumn{1}{c}{} & \multicolumn{1}{c}{0.99} & \multicolumn{1}{c}{1.00} & \multicolumn{1}{c}{1.00} & \multicolumn{1}{c}{1.26} & \multicolumn{1}{c}{} & \multicolumn{1}{c}{0.93} & \multicolumn{1}{c}{0.93} & \multicolumn{1}{c}{0.95} & \multicolumn{1}{c}{0.97} \\
		unst  & \multicolumn{1}{c}{3.4} & \multicolumn{1}{c}{0.4} & \multicolumn{1}{c}{-8.5} & \multicolumn{1}{c}{-6.1} & \multicolumn{1}{c}{} & \multicolumn{1}{c}{1.01} & \multicolumn{1}{c}{1.01} & \multicolumn{1}{c}{1.01} & \multicolumn{1}{c}{1.26} & \multicolumn{1}{c}{} & \multicolumn{1}{c}{0.95} & \multicolumn{1}{c}{0.95} & \multicolumn{1}{c}{0.97} & \multicolumn{1}{c}{0.95} \\
		unif  & \multicolumn{1}{c}{-0.1} & \multicolumn{1}{c}{2.0} & \multicolumn{1}{c}{3.2} & \multicolumn{1}{c}{1.6} & \multicolumn{1}{c}{} & \multicolumn{1}{c}{1.00} & \multicolumn{1}{c}{1.00} & \multicolumn{1}{c}{1.00} & \multicolumn{1}{c}{1.26} & \multicolumn{1}{c}{} & \multicolumn{1}{c}{0.94} & \multicolumn{1}{c}{0.94} & \multicolumn{1}{c}{0.94} & \multicolumn{1}{c}{0.95} \\
\end{tabular}%
\end{table}%

\begin{table}[htbp]
	\centering
	\scriptsize
	\caption{Evaluation criteria for misspecified models. Results for $n=600$ and $S=1000$ simulations.}\label{dois_n600_cor}
	\begin{tabular}{rrrrrrrrrrrrrrr}
		\hline
		& \multicolumn{4}{c}{Relative Bias} &       & \multicolumn{4}{c}{Relative Efficiency} &       & \multicolumn{4}{c}{Empirical Coverage} \\
		\cline{2-5}\cline{7-10}\cline{12-15}    Structure & $\beta_{01}$ & $\beta_{02}$ & $X$     & $Z$     &       & $\beta_{01}$ & $\beta_{02}$ & $X$     & $Z$     &       & $\beta_{01}$ & $\beta_{02}$ & $X$     & $Z$ \\
		\hline
		& \multicolumn{14}{c}{Available} \\
		ind   & \multicolumn{1}{c}{119.9} & \multicolumn{1}{c}{-31.8} & \multicolumn{1}{c}{-22.5} & \multicolumn{1}{c}{61.8} & \multicolumn{1}{c}{} & \multicolumn{1}{c}{1.00} & \multicolumn{1}{c}{1.00} & \multicolumn{1}{c}{1.00} & \multicolumn{1}{c}{1.00} & \multicolumn{1}{c}{} & \multicolumn{1}{c}{0.01} & \multicolumn{1}{c}{0.08} & \multicolumn{1}{c}{0.92} & \multicolumn{1}{c}{0.50} \\
		exch  & \multicolumn{1}{c}{95.2} & \multicolumn{1}{c}{-24.9} & \multicolumn{1}{c}{-25.1} & \multicolumn{1}{c}{30.8} & \multicolumn{1}{c}{} & \multicolumn{1}{c}{1.02} & \multicolumn{1}{c}{1.01} & \multicolumn{1}{c}{1.01} & \multicolumn{1}{c}{1.21} & \multicolumn{1}{c}{} & \multicolumn{1}{c}{0.08} & \multicolumn{1}{c}{0.27} & \multicolumn{1}{c}{0.89} & \multicolumn{1}{c}{0.76} \\
		unst  & \multicolumn{1}{c}{94.3} & \multicolumn{1}{c}{-26.2} & \multicolumn{1}{c}{-27.9} & \multicolumn{1}{c}{29.8} & \multicolumn{1}{c}{} & \multicolumn{1}{c}{1.03} & \multicolumn{1}{c}{1.02} & \multicolumn{1}{c}{1.02} & \multicolumn{1}{c}{1.21} & \multicolumn{1}{c}{} & \multicolumn{1}{c}{0.09} & \multicolumn{1}{c}{0.20} & \multicolumn{1}{c}{0.88} & \multicolumn{1}{c}{0.79} \\
		unif  & \multicolumn{1}{c}{93.5} & \multicolumn{1}{c}{-24.5} & \multicolumn{1}{c}{-28.5} & \multicolumn{1}{c}{25.3} & \multicolumn{1}{c}{} & \multicolumn{1}{c}{1.02} & \multicolumn{1}{c}{1.00} & \multicolumn{1}{c}{1.01} & \multicolumn{1}{c}{1.24} & \multicolumn{1}{c}{} & \multicolumn{1}{c}{0.08} & \multicolumn{1}{c}{0.26} & \multicolumn{1}{c}{0.90} & \multicolumn{1}{c}{0.82} \\
\end{tabular}%
\end{table}%

\begin{table}[htbp]
	\centering
	\scriptsize
	\caption{Evaluation criteria for correctly specified models. Results for $n=600$ and $S=1000$ simulations.}\label{dois_n600_cor}
	\begin{tabular}{rrrrrrrrrrrrrrr}
		\hline
		& \multicolumn{4}{c}{Relative Bias} &       & \multicolumn{4}{c}{Relative Efficiency} &       & \multicolumn{4}{c}{Empirical Coverage} \\
		\cline{2-5}\cline{7-10}\cline{12-15}    Structure & $\beta_{01}$ & $\beta_{02}$ & $X$     & $Z$     &       & $\beta_{01}$ & $\beta_{02}$ & $X$     & $Z$     &       & $\beta_{01}$ & $\beta_{02}$ & $X$    & $Z$\\
		\hline
		& \multicolumn{14}{c}{Complete} \\
		ind   & \multicolumn{1}{c}{1.0} & \multicolumn{1}{c}{-0.2} & \multicolumn{1}{c}{-1.0} & \multicolumn{1}{c}{0.3} & \multicolumn{1}{c}{} & \multicolumn{1}{c}{1.00} & \multicolumn{1}{c}{1.00} & \multicolumn{1}{c}{1.00} & \multicolumn{1}{c}{1.00} & \multicolumn{1}{c}{} & \multicolumn{1}{c}{0.96} & \multicolumn{1}{c}{0.96} & \multicolumn{1}{c}{0.95} & \multicolumn{1}{c}{0.93} \\
		exch  & \multicolumn{1}{c}{1.2} & \multicolumn{1}{c}{-0.1} & \multicolumn{1}{c}{0.1} & \multicolumn{1}{c}{0.2} & \multicolumn{1}{c}{} & \multicolumn{1}{c}{1.00} & \multicolumn{1}{c}{1.00} & \multicolumn{1}{c}{1.00} & \multicolumn{1}{c}{1.26} & \multicolumn{1}{c}{} & \multicolumn{1}{c}{0.97} & \multicolumn{1}{c}{0.96} & \multicolumn{1}{c}{0.96} & \multicolumn{1}{c}{0.95} \\
		unst  & \multicolumn{1}{c}{0.2} & \multicolumn{1}{c}{-0.2} & \multicolumn{1}{c}{-1.0} & \multicolumn{1}{c}{0.3} & \multicolumn{1}{c}{} & \multicolumn{1}{c}{1.00} & \multicolumn{1}{c}{1.00} & \multicolumn{1}{c}{1.00} & \multicolumn{1}{c}{1.26} & \multicolumn{1}{c}{} & \multicolumn{1}{c}{0.96} & \multicolumn{1}{c}{0.95} & \multicolumn{1}{c}{0.96} & \multicolumn{1}{c}{0.95} \\
		unif  & \multicolumn{1}{c}{0.9} & \multicolumn{1}{c}{-0.2} & \multicolumn{1}{c}{-0.9} & \multicolumn{1}{c}{0.2} & \multicolumn{1}{c}{} & \multicolumn{1}{c}{1.00} & \multicolumn{1}{c}{1.01} & \multicolumn{1}{c}{1.00} & \multicolumn{1}{c}{1.28} & \multicolumn{1}{c}{} & \multicolumn{1}{c}{0.96} & \multicolumn{1}{c}{0.95} & \multicolumn{1}{c}{0.95} & \multicolumn{1}{c}{0.94} \\
\end{tabular}%
\end{table}%

\newpage

\begin{table}[htbp]
	\centering
	\caption{Convergence Rate (CR) for the Simulation Study}\label{dois_converge}
	\begin{tabular}{rcccc}
		\hline
		& \multicolumn{4}{c}{sample size} \\
		\cline{2-5}          & 50    & 150   & 300   & 600 \\
		\cline{1-5}    ind   & 0.97  & 1.00  & 1.00  & 1.00 \\
		exch  & 0.59  & 0.89  & 0.96  & 1.00 \\
		unst  & 0.08  & 0.59  & 0.89  & 0.99 \\
		unif  & 0.97  & 1.00  & 1.00  & 1.00 \\
		cat.exch & 0.97  & 1.00  & 1.00  & 1.00 \\
		time.exch & 0.96  & 1.00  & 1.00  & 1.00 \\
		RC    & 0.67  & 0.99  & 1.00  & 1.00 \\
		\hline
	\end{tabular}%
\end{table}%

Table \ref{dois_converge} shows the convergence rate obtained from the simulation results for seven association structures and four sample sizes. The convergence rate for the working association structure (C) was calculated as 
$$
CR^C=\frac{1000}{\mbox{total number of simulations}}.
$$
The local odds ratio parametrization presents a clear advantage over the correlation coefficient in terms of convergence issues, specially for unstructured association matrices and small sample sizes. 

\newpage
\subsection{Additional Real Data Results}
\begin{table}[htbp]
	\centering
	\caption{Results for Rheumatic Mitral Stenosis study under exchangeability, time exchangeability and RC association structures}\label{dois_tab4}
	\begin{tabular}{rcccrcccrccc}
		\hline
		& \multicolumn{3}{c}{Exchangeable} &       & \multicolumn{3}{c}{Time Exchangeability} &       & \multicolumn{3}{c}{RC} \\
		\cline{2-4}\cline{6-8}\cline{10-12}    Parameter & Est.  & SE    & p     &       & Est.  & SE    & p     &       & Est.  & SE    & p \\
		\cline{1-4}\cline{6-8}\cline{10-12}    \textbf{} & \multicolumn{11}{c}{Available} \\
		$\beta_{01}$ & -1.064 & 0.782 & 0.174 &       & -0.909 & 0.730 & 0.213 &       & -0.881 & 0.744 & 0.236 \\
		$\beta_{02}$ & 0.721 & 0.776 & 0.353 &       & 0.890 & 0.724 & 0.219 &       & 0.924 & 0.738 & 0.210 \\
		Success & 0.680 & 0.397 & 0.086 &       & 0.626 & 0.387 & 0.105 &       & 0.641 & 0.391 & 0.101 \\
		Total Score & -0.108 & 0.107 & 0.313 &       & -0.130 & 0.101 & 0.197 &       & -0.137 & 0.102 & 0.179 \\
		Cn    & 0.468 & 0.339 & 0.168 &       & 0.477 & 0.314 & 0.129 &       & 0.496 & 0.317 & 0.118 \\
		time=2 & 1.083 & 0.398 & 0.007 &       & 1.022 & 0.383 & 0.008 &       & 1.081 & 0.390 & 0.006 \\
		time=3 & 1.208 & 0.402 & 0.003 &       & 1.120 & 0.380 & 0.003 &       & 1.214 & 0.392 & 0.002 \\
		& \multicolumn{11}{c}{WGEE} \\
		$\beta_{01}$ & -1178 & 0.807 & 0.144 &       & -0.934 & 0.756 & 0.217 &       & -0.892 & 0.771 & 0.248 \\
		$\beta_{02}$ & 0.611 & 0.797 & 0.443 &       & 0.848 & 0.741 & 0.252 &       & 0.898 & 0.757 & 0.236 \\
		Success & 0.744 & 0.410 & 0.070 &       & 0.623 & 0.400 & 0.119 &       & 0.645 & 0.404 & 0.111 \\
		Total Score & -0.101 & 0.109 & 0.355 &       & -0.127 & 0.103 & 0.216 &       & -0.136 & 0.105 & 0.195 \\
		Cn    & 0.490 & 0.351 & 0.162 &       & 0.433 & 0.322 & 0.179 &       & 0.449 & 0.325 & 0.167 \\
		time=2 & 1.037 & 0.414 & 0.012 &       & 1.013 & 0.398 & 0.011 &       & 1.060 & 0.404 & 0.009 \\
		time=3 & 1.395 & 0.418 & 0.001 &       & 1.363 & 0.405 & 0.001 &       & 1.437 & 0.415 & 0.001 \\
		& \multicolumn{11}{c}{MIGEE} \\
		$\beta_{01}$ & -1.100 & 0.618 & 0.075 &       & -0.865 & 0.573 & 0.131 &       & -0.908 & 0.567 & 0.109 \\
		$\beta_{02}$ & 0.743 & 0.621 & 0.231 &       & 0.973 & 0.575 & 0.091 &       & 0.934 & 0.572 & 0.102 \\
		Success & 1.079  & 0.334 & 0.001 &       & 0.834 & 0.322 & 0.010 &       & 0.887 & 0.324 & 0.006 \\
		Total Score & -0.098 & 0.083 & 0.239 &       & -0.131 & 0.078 & 0.094 &       & -0.131 & 0.078 & 0.093 \\
		Cn    & 0.237 & 0.280 & 0.397 &       & 0.281 & 0.265 & 0.288 &       & 0.350 & 0.272 & 0.198 \\
		time=2 & 1.095  & 0.317 & 0.001 &       & 1.075 & 0.320 & 0.001 &       & 1.033 & 0.323 & 0.001 \\
		time=3 & 1.071  & 0.316 & 0.001 &       & 1.091 & 0.308 & 0.000 &       & 1.072 & 0.313 & 0.001 \\
		& \multicolumn{11}{c}{DRGEE} \\
		$\beta_{01}$ & -1.199 & 0.781 & 0.125 &       & -0.896 & 0.692 & 0.195 &       & -0.852 & 0.700 & 0.224 \\
		$\beta_{02}$ & 0.591 & 0.785 & 0.451 &       & 0.888 & 0.686 & 0.195 &       & 0.934 & 0.696 & 0.180 \\
		Success & 0.808 & 0.380 & 0.034 &       & 0.727 & 0.368 & 0.048 &       & 0.748 & 0.371 & 0.044 \\
		Total Score & -0.097 & 0.108 & 0.368 &       & -0.137 & 0.094 & 0.148 &       & -0.143 & 0.096 & 0.135 \\
		Cn    & 0.475 & 0.373 & 0.202 &       & 0.490 & 0.326 & 0.133 &       & 0.495 & 0.330 & 0.134 \\
		time=2 & 0.937 & 0.630 & 0.137 &       & 0.985 & 0.534 & 0.065 &       & 0.983 & 0.540 & 0.069 \\
		time=3 & 1.203 & 0.601 & 0.046 &       & 1.189 & 0.498 & 0.017 &       & 1.208 & 0.507 & 0.017 \\
		\hline
	\end{tabular}%
	\label{tab:addlabel}%
\end{table}%

\end{document}